\documentclass[aps,prb,groupaddress,twocolumn,showpacs]{revtex4-1} 

\usepackage[colorlinks=true, linkcolor=blue, citecolor=blue, urlcolor=blue]{hyperref}
\usepackage{amsmath,amssymb,mathrsfs}
\usepackage{latexsym}
\usepackage{graphicx} 
\usepackage{epstopdf}
\usepackage{graphicx,epstopdf,color}
\usepackage{amsfonts}
\usepackage{bm}
\usepackage{bbm}
\usepackage{multirow}
\usepackage{rotating}

\graphicspath{./}
\definecolor{forestgreen}{rgb}{0.13, 0.60, 0.13}

\def\cf{\emph{c.f.}\ }
\def\ie{\emph{i.e.},\ }
\def\eg{\emph{e.g.}\ }

\newcommand{\bs}[1]{\boldsymbol{#1}}

\newcommand{\up}{\uparrow}
\newcommand{\dw}{\downarrow}
\newcommand{\pd}{{\phantom{\dagger}}}

\newcommand{\hel}{\zeta}
\newcommand{\evz}{\theta}

\begin{document}

\title{Spin-orbit coupled superconductivity:\\
Rashba--Hubbard model on the square lattice}

\author{Sebastian Wolf}
\author{Stephan Rachel}
\affiliation{School of Physics, University of Melbourne, Parkville, VIC 3010, Australia}
 \pagestyle{plain}

\begin{abstract}
The weak-coupling renormalization group method is an asymptotically exact method to find superconducting instabilities of a lattice model of correlated electrons. Here we extend it to spin-orbit coupled lattice systems and study the emerging superconducting phases of the Rashba--Hubbard model. Since Rashba type spin-orbit coupling breaks inversion and spin symmetry, the arising superconducting phases may be a mixture of spin-singlet and spin-triplet states. We study the two-dimensional square lattice as a paradigm and discuss the symmetry properties of the arising spin-orbit coupled superconducting states including helical spin-triplet superconductivity. We also discuss how to best deal with split energy bands within a method which restricts paired electrons to momenta on the Fermi surface.
\end{abstract}

%\date{\today}

\maketitle

\section{Introduction} 
\label{sec:Intro}
The discovery of quantum spin Hall and topological insulators\,\cite{hasan_colloquium_2010,qi_topological_2011} as well as Weyl and Dirac semimetals\,\cite{yan_topological_2017} has brought spin-orbit coupling (SOC) into the spotlight of condensed matter research. SOC constitutes the most crucial ingredient for this rich and diverse family of materials. Topological superconductivity\,\cite{sato_topological_2017,kallin_chiral_2016,tanaka-12jpsp011013,smidman-17rpp036501} can also be stabilized in spin-orbit coupled metals via proximity induced superconductivity\,\cite{mourik_signatures_2012,nadj-perge_observation_2014,palacio-morales_atomic-scale_2019}; alternatively, it arises as an intrinsic many-body instability in correlated electron systems. The latter requires odd-parity spin-triplet pairing, as realized in the chiral $p$-wave state with its Majorana zero modes at defects or sample boundaries\,\cite{read_paired_2000,ivanov_non-abelian_2001}. SOC is not a requirement for such exotic instabilities, but today it is apparent that SOC is beneficial for triplet pairing\,\cite{yanase-08jpsp124711,yokoyama-07prb172511}. Most of the few candidate materials for triplet superconductivity contain heavy elements and thus significant SOC, \eg Sr$_2$RuO$_4$\,\cite{mackenzie-03rmp657,pustogow-19n72}, CePt$_{3}$Si\,\cite{bauer-04prl027003,yanase-08jpsp124711}, Cu$_x$BiSe$_2$\,\cite{hor-10prl057001,sasaki-11prl217001}, and most recently UTe$_2$\,\cite{ran-19s684}. Nonetheless, the role of spin-orbit coupled superconductivity is surprisingly underrepresented in the literature.

Quite generally, we distinguish between different types of SOC in solids. $\vec L\cdot\vec S$ corresponds to centrosymmetric SOC; it preserves not only inversion symmetry but also the spin degeneracy of the SOC-free systems (while spin symmetry is broken). Rashba and Dresselhaus terms are stemming from the breaking of inversion symmetry and correspond thus to non-centrosymmetric SOC. They break the spin degeneracy. Material examples of the latter type include the doped Weyl semimetals WTe$_2$, MoTe$_2$\,\cite{pan_pressure-driven_2015,qi_superconductivity_2016}, and YPtBi\,\cite{butch_superconductivity_2011,savary-17prb214514,kim_beyond_2018}, as well as the heavy fermion compound CePt$_3$Si\,\cite{smidman-17rpp036501,bauer-04prl027003}; 
these materials are unconventional superconductors at sufficiently low temperatures, and CePt$_3$Si has even been claimed to realize triplet pairing\,\cite{bauer-04prl027003,yanase-08jpsp124711,smidman-17rpp036501}.

There are several notable works in the literature using various methods to study the effect of Rashba SOC to correlated electrons
%Rashba-Hubbard model, \ie free electrons in the presence of Rashba SOC and an interaction term, 
and the resulting superconducting instabilities. In particular, these include a ``Shankar RG'' approach\,\cite{shankar_renormalization-group_1994} for a continuum model\,\cite{vafek_spin-orbit_2011,wang_unconventional_2014}, random phase approximation (RPA) studies on the square lattice\,\cite{shigeta_superconducting_2013,greco_mechanism_2018,ghadimi-19prb115122} as well as work tailored for the materials CePt$_{3}$Si and Li$_{2}$Pd$_{x}$Pt$_{3-x}$B\,\cite{bittner_leggett_2015}. Moreover, there are several works using Bardeen--Cooper--Schrieffer (BCS) theory for continuum systems\,\cite{gorkov_superconducting_2001} and specifically for CePt$_{3}$Si\,\cite{sergienko_order_2004,frigeri_spin_2004,frigeri_superconductivity_2004,frigeri_characterization_2006}. The common conclusion from these works is that the breaking of inversion symmetry due to Rashba SOC causes mixed singlet-triplet superconducting states to appear. Furthermore, {\it strong} SOC suppresses chiral states, \ie the only topologically nontrivial states that may arise in the superconducting condensate are helical ones characterized by a $\mathbb{Z}_{2}$ invariant. The mixing of singlet and triplet states can be understood as follows: when spin symmetry in a superconductor is broken, parity still remains well defined. That is, the momentum-part of the superconducting order parameter is either an even, \ie symmetric function in momentum (even parity, spin singlet) or an odd, \ie antisymmetric function in momentum (odd parity, spin triplet). Once inversion symmetry is broken parity is no longer a good quantum number and the resulting states can be mixtures of even- and odd-parity functions, thus allowing for singlet-triplet mixtures.

In this paper, we study the superconducting states on the square lattice as a paradigm for 2D systems in the presence of 
non-centrosymmetric, \ie Rashba, SOC.
%strong inversion symmetry breaking spin-orbit coupling (SOC). 
Considering infinitesimal repulsive  interactions allows us to calculate the arising superconducting instabilities {\it exactly} using the weak coupling renormalization group (WCRG) approach\,\cite{raghu_superconductivity_2010,raghu_effects_2012,wolf_unconventional_2018,cho-13prb064505,platt-16prb214515,cho-15prb134514,scaffidi-14prb220510,roising-18prb224515}.
We choose this method since it poses several advantages: firstly, the results are asymptotically exact in the limit of vanishing interaction. Secondly, compared to mean field methods, we obtain the resulting superconducting instabilities in an unbiased way, \ie without assumptions on their properties. This is particularly advantageous here, since breaking of inversion symmetry\,\cite{smidman-17rpp036501} may lead to mixed singlet and triplet states, which yields a large variety of possible pairing states. Thirdly, compared to functional renormalization group methods, which is widely accepted to  perform well at stronger interactions\,\cite{metzner-12rmp299,platt-13ap453}, the WCRG is computationally very efficient. Especially the doubling of the number of bands due to the Rashba SOC demands a very efficient method to make the numerical integrations feasible.

The paper is organized as follows: In  Sec.\,\ref{sec:method}, we give a description of how Rashba SOC is implemented in the WCRG framework and what kind of mixed superconducting instabilities one can generally expect to find on the square lattice. In Sec.\,\ref{sec:results} we study the competing superconducting pairing channels of the square lattice Rashba-Hubbard model, followed by a thorough discussion on how to interpret the results and an outlook in Sec.\,\ref{sec:discussion}. Sec.\,\ref{sec:conclusion} contains the paper's conclusion, followed by four appendices with further details.

%%%%%%%%%%%%%%%%%%%%%%%%%%%%%%%%%%%%%%%%%%%%%%%%%%%%%%%%%%%%%%%%%
%
%                               M O D E L    A N D    M E T H O D
%
%%%%%%%%%%%%%%%%%%%%%%%%%%%%%%%%%%%%%%%%%%%%%%%%%%%%%%%%%%%%%%%%%
\section{Model and Method}
\label{sec:method}
\subsection{Hamiltonian and Bandstructure}
\label{subsec:Hamilton_BS}
We consider the Hubbard Hamiltonian including SOC in the form of an additive Rashba term, $H_{R}$. For simplicity, we restrict ourselves to systems where there is only one orbital per unit cell. The Hamiltonian can be written as
\begin{equation}
\label{eq:total_general_ham}
 H=H_{0}+H_{\rm int}+H_{R}
 \end{equation}
 with
 \begin{align}
\label{eq:ham_hopping_term}
 H_{0}&=\sum_{i,j}\sum_{\sigma}
            t_{ij}c_{i\sigma}^{\dagger}c_{j\sigma}^{\pd}\ ,\\
\label{eq:ham_interaction_term}
 H_{\rm int}&=\sum_{i}\sum_{\sigma,\sigma'}
            U_{0,\sigma\sigma'}c_{i\sigma}^{\dagger}c_{i\sigma'}^{\dagger}c_{i\sigma'}^{\pd}c_{i\sigma}^{\pd}\ ,\\
\label{eq:ham_Rashba_SOC_term}
 H_{R}&=\sum_{i,j}\sum_{\sigma,\sigma'}
            i\,\alpha_{R}\,t_{ij}\,c_{i\sigma}^{\dagger}c_{j\sigma'}^{\pd}\left(\vec{\sigma}\times\vec{r}_{ij}\right)_{z,\sigma\sigma'}\ ,
\end{align}
where $c_{i\sigma}$ is the annihilation operator of an electron with spin $\sigma$ at lattice site $i$. The hopping amplitude is denoted by $t_{ij}$; throughout the paper we only consider $t_{ij}\equiv t_1$ ($t_{ij}\equiv t_2$) for $i$ and $j$ being nearest (next-nearest) neighbors. $U_{0}$ denotes the onsite interaction strength, $\alpha_{R}$ the dimensionless strength of  Rashba SOC, $\vec{\sigma}$ is the vector of Pauli matrices and $\vec{r}_{ij}$ measures the distance between sites $i$ and $j$. The Fourier transform of the non-interacting part,
\begin{equation}
\label{eq:bloch}
 H_{0}+H_{R}=\sum_{k}   (c_{k\up}^{\dagger},c_{k\dw}^{\dagger})\,
                        \hat{h}(k)
                        \begin{pmatrix}
                            c_{k\up}^{\pd} \\
                            c_{k\dw}^{\pd}
                        \end{pmatrix},
\end{equation}
yields the Bloch matrix $\hat{h}$, 
\begin{equation}
\label{eq:bloch_general}
    \hat{h}=\gamma_{\nu}\sigma^{\nu},\hspace{2mm}\nu\in\{0,x,y,z\}\ ,
\end{equation}
where $k$ denotes the momentum vector $(k_{x},k_{y},k_{z})^{T}$.
In Eq.\,\eqref{eq:bloch_general} we used the 4-vector notation and $\gamma_{0}(k)$ is the energy spectrum without SOC.

The full energy spectrum $E(k,\hel)$ of the non-interacting system is readily obtained by diagonalizing $\hat{h}$ via unitary transformation,
\begin{eqnarray}
\label{eq:total_spectrum}
 &\xi(k)=\hat{U}^{\dagger}(k)\,\hat{h}(k)\,\hat{U}(k)=\begin{pmatrix}
         E(k,-) & 0 \\
         0 & E(k,+)
        \end{pmatrix},&\\[10pt]
& E(k,\hel)=\gamma_{0}(k)+\hel|\vec{\gamma}(k)|\ .
\end{eqnarray}
The unitary matrix $\hat{U}(k)$ has the eigenvectors, $\vec{v}(k,\hel)$, of $\hat{h}(k)$ as column vectors, which are given by
\begin{equation}
 \vec{v}(k,\hel)=\frac{1}{\sqrt{2}}\begin{pmatrix}
             \evz_{\hel}(k) \\
             \hel e^{i\phi_{k}}\evz_{-\hel}(k)
            \end{pmatrix}
\end{equation}
            with
\begin{equation}
 \evz_{\hel}(k)=\sqrt{1+\hel\gamma_{z}/|\vec{\gamma}|}\ ,\hspace{4mm}
 e^{i\phi_{k}}=\frac{\gamma_{x}+i\gamma_{y}}{\sqrt{\gamma_{x}^{2}+\gamma_{y}^{2}}}\ .
\end{equation}
The diagonalization effectively transforms from spin to helicity basis, with helicity quantum numbers $\hel=\pm1$. 

Until now the results are general for any $2\times2$ Bloch matrix, since any Hermitian matrix can be written as Eq.\,\eqref{eq:bloch_general}. Considering Rashba SOC, Eqs.\,\eqref{eq:ham_Rashba_SOC_term} and \eqref{eq:bloch_general} result in
\begin{align}
\label{eq:gamma_vector}
 \vec{\gamma}(k)\equiv&(\gamma_{x},\gamma_{y},\gamma_{z})^{T}=-\alpha_{R}(\hat{z}\times\nabla)\gamma_{0}(k)
\end{align}
and
\begin{align}
 \hat{h}=&\begin{pmatrix}
          \gamma_{0} & \alpha_{R}\Big[\frac{\partial\gamma_{0}}{\partial k_{y}}+i\frac{\partial\gamma_{0}}{\partial k_{x}}\Big] \\
          \alpha_{R}\Big[\frac{\partial\gamma_{0}}{\partial k_{y}}-i\frac{\partial\gamma_{0}}{\partial k_{x}}\Big] & \gamma_{0}
         \end{pmatrix}.
\end{align}
In particular, it follows that $\gamma_{z}=0$ leading to $\evz_{\hel}=1$ which is a consequence of time-reversal symmetry and invariance under $\pi$ rotations around $\hat z$ (see App.\,\ref{sec:appD}).

\subsection{Weak Coupling RG}
\label{subsec:WCRG}
We employ the WCRG method to find the leading superconducting instability in the Hubbard model in the presence of SOC. This method has been discussed in great detail for fermions without spin mixing before\,\cite{raghu_superconductivity_2010,raghu_effects_2012,wolf_unconventional_2018}, so only a brief summary is given here, but we point out important differences which arise when one includes spin mixing effects such as spin-orbit coupling. Note that in the following we limit ourselves to systems where the spin degeneracy has been lifted, which is achieved for example by Rashba SOC. Furthermore, we assume that time reversal symmetry is conserved. This offers a major simplification: we do not need to consider the spin degree of freedom for the electron states, as this is fixed already by the weak coupling nature of the method. For example, if we consider a scattering process of two electrons that contributes to the Cooper channel in the weak coupling regime, their initial momenta have to be opposite and both on the Fermi surface, \ie the electrons have to be initially in the same band. Since each band has only one spin polarization per momentum, there is no freedom in choosing the spin of the scattering particles, it is dictated by the helicity of the band:
\begin{equation}
 \begin{bmatrix}
  k,\sigma\\
  -k,\sigma'
 \end{bmatrix}\hspace{3mm}\Rightarrow\hspace{3mm}
 \begin{bmatrix}
  k,\hel\\
  -k,\hel
 \end{bmatrix}.
\end{equation}

In the following, we use the short notation for momentum and helicity,
\begin{equation}
 1\equiv k_{1},\hel_{1}\ ,\hspace{6mm}\bar{1}\equiv-k_{1},\hel_{1}\ .
\end{equation}

\begin{figure}[t]
 \includegraphics[width=0.9\columnwidth]{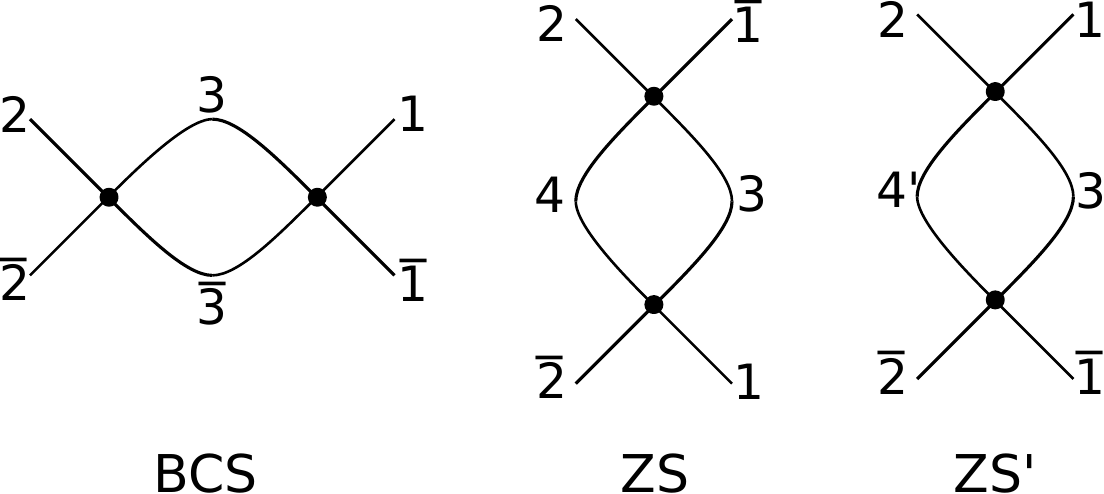}
 \caption{\label{fig:Feynman}Feynman diagrams representing all topologically distinct second order contributions to the two-particle vertex function for spinful fermions\,\cite{shankar_renormalization-group_1994}. For details see main text and App.\,\ref{sec:appB}.}
\end{figure}

The important quantity for calculating the superconducting instabilities in the WCRG method is the two-particle vertex $\Gamma$ in the Cooper channel.
Since the method explicitly demands weak coupling, we expand $\Gamma$ in orders of the local electron-electron interaction, $U_{0}$, up to second order,
\begin{align}
 \Gamma(2,1)=&\,U_{0}\Gamma^{(1)}(2,1)+U_{0}^{2}\Gamma^{(2)}(2,1)+\dots
\end{align}
The first order, $\Gamma^{(1)}$, is given by\,\cite{vafek_spin-orbit_2011}
\begin{align}
\label{eq:Gamma1}
 \Gamma^{(1)}(2,1)=&\,M(2\bar{2}\bar{1}1)=\hel_{2}\hel_{1}e^{i(\phi_{k_{2}}-\phi_{k_{1}})}\ ,
\end{align}
where the last equality stems from the specific system outlined above, and the factor $M$ reads
\begin{align}
 M(4321)=&\,
  \langle\vec{v}(4),\vec{v}(1)\rangle\langle\vec{v}(3),\vec{v}(2)\rangle\nonumber\\[5pt]
 &-\,\langle\vec{v}(4),\vec{v}(2)\rangle\langle\vec{v}(3),\vec{v}(1)\rangle\ .
\end{align}
The complex scalar product is herein denoted as $\langle\cdot,\cdot\rangle$. 
The second order, $\Gamma^{(2)}$, splits into three topologically distinct parts\,\cite{shankar_renormalization-group_1994,vafek_spin-orbit_2011},
\begin{equation}
\label{eq:Gamma2_tot}
 \Gamma^{(2)}=\,\frac{1}{2}\Gamma_{\rm BCS}+\Gamma_{\rm ZS}+\Gamma_{\rm ZS'}\ .
 \end{equation}
 The first contribution is referred to as the BCS diagram and the other two contributions to zero sound (ZS) diagrams\,\cite{shankar_renormalization-group_1994}; the vertices are given by
 \begin{align}
 \Gamma_{\rm BCS}(2,1)=&-\int_{3}M(2\bar{2}\bar{3}3)M(3\bar{3}\bar{1}1)X_{\rm pp}(3)\nonumber\\
\label{eq:Gamma2_BCS}
                 =&-\hel_{1}\hel_{2}e^{i(\phi_{k_{2}}-\phi_{k_{1}})}\rho\ln\Big(\frac{A}{\epsilon}\Big)\ ,\\[10pt]
\label{eq:Gamma2_ZS}
 \Gamma_{\rm ZS}(2,1)=&\int_{3'}M(\bar{2}431)M(234\bar{1})X_{\rm ph}(3,4)\ ,\\[10pt]
 \Gamma_{\rm ZS'}(2,1)=&-\int_{3'}M(24'31)M(\bar{2}34'\bar{1})X_{\rm ph}(3,4')\nonumber\\
\label{eq:Gamma2_ZS'}
                =&-\Gamma_{\rm ZS}(\bar{2},1)\ ,
\end{align}
for which the respective Feynman diagrams are shown in Fig.\,\ref{fig:Feynman}. Here, $\epsilon$ is an energy cut-off in the integral and $A$ is a system parameter which merely depends on the bandstructure,
\begin{equation}
\ln{(A)} = \int^{E_{\rm max}} \frac{dE}{2E}\frac{\rho(E)}{\rho} + \int^{E_{\rm min}} \frac{dE}{2E}\frac{\rho(E)}{\rho}   \ ,
\end{equation}
with the density of states $\rho(E)$, $\rho\equiv\rho(0)$, and $E_{\rm min/max}$ the minimum/maximum of the bandstructure. Note that we are showing here the diagrams used in the review by Shankar\,\cite{shankar_renormalization-group_1994}. 
We note that these diagrams seem to be different from those previously introduced in Ref.\,\onlinecite{raghu_superconductivity_2010}; the relationship between the diagrams in Fig.\,\ref{fig:Feynman} and in Ref.\,\onlinecite{raghu_superconductivity_2010} is explained in App.\,\ref{sec:appB}.

%
%%%%%%%%%%%%%%%%%%%%%%%%%%%%%%%%%%%%%%%%%%%%%%%%%%%%%%%%%%%%%%
%           F I G. 1
%%%%%%%%%%%%%%%%%%%%%%%%%%%%%%%%%%%%%%%%%%%%%%%%%%%%%%%%%%%%%%
%%%%%%%%%%%%%%%%%%%%%%%%%%%%%%%%%%%%%%%%%%%%%%%%%%%%%%%%%%%%%%
%
Momentum conservation yields $4\equiv k_{1}+k_{2}+k_{3},\hel_{4}$ and $4'\equiv k_{1}-k_{2}+k_{3},\hel_{4}$. 
The short notations for the integrals are
\begin{align}
\label{eq:short_int_1}
 \int_{3}\equiv\sum_{\hel_{3}}\int\frac{\text{d}^{2}k_{3}}{(2\pi)^{2}}\ ,\hspace{4mm} \int_{3'}\equiv\sum_{\hel_{4}}\int_{3}\ .
\end{align}
$X_{\rm ph}$ ($X_{\rm pp}$) denotes the integrands of the static particle-hole susceptibility $\chi_{\rm ph}$ (static particle-particle susceptibility $\chi_{\rm pp}$), 
\begin{align}
 \chi_{\rm ph}(\vec{k}_{4}-\vec{k}_{3})=&\,-\int_{3'}X_{\rm ph}(3,4)\ ,\\ %=-\int_{3'}\frac{ f(E(3)) - f(E(4))}{E(3)-E(4)},\\
 X_{\rm ph}(3,4) =&\, \frac{ f(E(3)) - f(E(4))}{E(3)-E(4)}\ ,\\[10pt]
 \chi_{\rm pp}(\epsilon)=&\,\int_{3,|E|>\epsilon}X_{\rm pp}(3)\ ,\\
 X_{\rm pp}(3)=&\,-\frac{1-2 f(E(3))}{2E(3)}\ .
\end{align}
Here $f(E)$ is the Fermi distribution.
$\Gamma_{\rm BCS}$ is the only diagramm in second order showing a logarithmic divergence. As derived in previous works\,\cite{raghu_superconductivity_2010,vafek_spin-orbit_2011,wang_unconventional_2014,wolf_unconventional_2018}, for repulsive interactions, $U_{0}>0$, the superconducting instabilities can be obtained from the non-divergent parts of the full vertex. This is seen with the help of the $\beta$-function of the RG method,
\begin{equation}
\label{eq:RG_flow}
\beta\left(\Gamma(2,1)\right)= \frac{\partial\Gamma(2,1)}{\partial\ln(\epsilon_{0}/\epsilon)}=-\int_{\hat{3}}\Gamma(2,\hat{3})\Gamma(\hat{3},1)\ ,
\end{equation}
where the hat denotes momenta at the Fermi level, $\epsilon_0$ the initial energy cut-off, and $\epsilon$ the lowered cut-off in the process of renormalization. The short notation for the integral is given by
\begin{equation}
\label{eq:FS_int_short}
 \int_{\hat{i}}\equiv\sum_{\hel_{i}}\rho_{\hel_{i}}\int_{\rm FS}\frac{\text{d}\hat{k}_{i}}{S_{\hel_{i},F}}\frac{\bar{v}_{\hel_{i},F}}{v_{F}(\hat{i})},
\end{equation}
where $v_{F}(\hat{i})$ denotes the Fermi velocity of band $\hel_{i}$ at momentum $\hat{k}_{i}$, $S_{\hel_{i},F}$ is the total length of the Fermi surface of band $\hel_{i}$, the integral runs over the Fermi surface, and
\begin{equation}
 \frac{1}{\bar{v}_{\hel_{i},F}}=\int_{\rm FS}\frac{d\hat{k}_{i}}{S_{\hel_{i},F}}\frac{1}{v_{F}(\hat{i})}.
\end{equation}

In second order in $U_{0}$, the non-diverging parts of $\Gamma$ are $\Gamma_{\rm ZS}$ and $\Gamma_{\rm ZS'}$. Thus, all we need to calculate is
\begin{align}
 \Gamma^{(2)}(2,1)\approx&\,\Gamma_{\rm ZS}(2,1)+\Gamma_{\rm ZS'}(2,1)\\[5pt]
 \label{eq:Gamma2_antisymm}
 =&\,\Gamma_{\rm ZS}(2,1)-\Gamma_{\rm ZS}(\bar{2},1)
\end{align}
where the second line is an explicit antisymmetrization. When dealing with fermions without spin mixing, as done in Refs.\,\onlinecite{raghu_superconductivity_2010,wolf_unconventional_2018}, we neglect the spin part of the full vertex. That is, $\Gamma$ can be symmetric (spin-singlet Cooper pairs) or antisymmetric (spin-triplet Cooper pairs), since we only work with the momentum-space part of $\Gamma$ and treat the spin-space part only implicitly. Here, however, we have to explicitly include spin because of the spin-orbit coupling, and the full fermionic vertex function has to be antisymmetric, which is reflected in Eq.\,\eqref{eq:Gamma2_antisymm}.

As first pointed out by Anderson\,\cite{anderson_theory_1959} and later by Sergienko and Curnoe\,\cite{sergienko_order_2004}, if spin is not a good quantum number we have to ensure that the two-particle scattering processes that are considered in $\Gamma$ are between time reversal partners.
The vertex, where for the incoming and outgoing electron pairs the two electrons are restricted to be time-reversal partners, will be called $\Gamma^{(T)}$, which differs from $\Gamma$ by a momentum and helicity dependant phase which is odd in momentum\,\cite{sergienko_order_2004},
\begin{eqnarray}
\label{eq:gamma_TR_phase}
& \Gamma(2,1)=e^{i\varphi(2,1)}\,\Gamma^{(T)}(2,1)\ ,&\\[8pt]
\label{eq:gamma_TR_phase_odd}
 &e^{i\varphi(\bar{2},1)}=e^{i\varphi(2,\bar{1})}=-e^{i\varphi(2,1)}=-e^{i\varphi(\bar{2},\bar{1})}\ ,&
 \end{eqnarray}
implying that $\Gamma^{(T)}$ must be even in momentum. 
In general, the total phase factor can be written as the product of two phase factors, each depending only on the momentum and helicity of the incoming and outgoing electrons, respectively, \ie
\begin{equation}
\label{eq:Gamma_TR}
 e^{i\varphi(2,1)}=t^{*}(2)\,t(1)\ .
\end{equation}

Computing the time reversal conjugate of the annihilation operators explicitly, \ie $T c^{\dagger}(1) = t(1)c^{\dagger}(\bar{1})$ [\cf App.\,\ref{sec:appC}], for the system discussed in Sec.\,\ref{subsec:Hamilton_BS} 
we obtain the phase factor as\,\cite{sergienko_order_2004}
\begin{equation}
\label{eq:TR_SOC_phase}
 t(1)=-\hel_{1}e^{-i\phi_{k_{1}}}\ .
\end{equation}
Eqs.\,\eqref{eq:Gamma1}, \eqref{eq:Gamma_TR}, and \eqref{eq:TR_SOC_phase} thus yield
\begin{align}
 \Gamma^{(1)(T)}(2,1)=1\ .
\end{align}
We thus find that the first order contribution only suppresses the isotropic $s$-wave superconductivity, as it happens for the single-orbital case without Rashba SOC\,\cite{raghu_superconductivity_2010,wolf_unconventional_2018}.

The effective interaction of the leading superconducting instability, $U_{\rm eff}=\rho\lambda_{\rm min}$, and the corresponding form factor of the superconducting order parameter, $\psi_{\rm min}$, are then obtained as follows\cite{raghu_superconductivity_2010}: First, we rescale $\Gamma$ to
\begin{align}
 g^{(T)}(\hat{2},\hat{1})&:=\tau(\hat{2})\Gamma^{(T)}(\hat{2},\hat{1})\tau(\hat{1})\ ,
\end{align}
where $\tau$ is obtained from Eq.\,\eqref{eq:FS_int_short} by
\begin{equation}
 \int_{\hat{i}}=\sum_{\hel_{i}}\int_{\rm FS}{\rm d}\hat{k}_{i}\,\tau^{2}(\hat{i})\ .
\end{equation}
Substituting an orthonormal eigenbasis of $g$, \ie
\begin{align}\label{eq:ev-equation}
 \sum_{\hel_{1}}\int_{\rm FS}{\rm d}\hat{k}_{1}\,g^{(T)}(\hat{2},\hat{1})\psi_{\nu}^{(T)}(\hat{1})&=\,\lambda_{\nu}\psi_{\nu}^{(T)}(\hat{2})\ ,\\
 \sum_{\hel_{1}}\int_{\rm FS}{\rm d}\hat{k}_{1}\,\big[\psi^{(T)}_{\nu}(\hat{1})\big]^{*}\psi^{(T)}_{\eta}(\hat{1})&=\delta_{\nu\eta}\ ,
\end{align}
into the flow equation, Eq.\,\eqref{eq:RG_flow}, yields that each eigenvalue of $g$ renormalizes independently, \ie
\begin{equation}
\label{eq:flow_lambda}
 \frac{\partial\lambda_{\nu}}{\partial\ln(\epsilon_{0}/\epsilon)}=-\lambda_{\nu}^{2}\ .
\end{equation}
Note that Eq.\,\eqref{eq:ev-equation} can also be understood as the linearized gap equation.
Thus, the most negative eigenvalue, $\lambda_{\rm\min}$, is the one that diverges first in the RG flow and, hence, serves as a measure for the critical temperature via
\begin{equation}
 T_{c}\sim e^{-1/|\lambda_{\rm min}|}=e^{-1/\rho |U_{\rm eff}|}\ .
\end{equation}

An analogous relationship to Eqs.\,\eqref{eq:gamma_TR_phase} and \eqref{eq:Gamma_TR} holds for the eigenvectors,
 \begin{equation}
 \psi_{\nu}(1)=\,t(1)\psi_{\nu}^{(T)}(1)\ .
\end{equation}
Note that each $\psi_{\nu}^{(T)}$ has to transform according to an \emph{even} irreducible representation (\ie the basis functions are even in momentum) of the symmetry group of the lattice, since $\Gamma^{(T)}$ is even in momentum.

The matrix of superconducting order parameters in spin space is obtained from the result in helicity space, $\psi(k,\hel)$, by the transformation
\begin{align}
 \hat{\Delta}(k)\equiv&\,\begin{pmatrix}
                    \psi(k,\up\up) & \psi(k,\up\dw) \\
                    \psi(k,\dw\up) & \psi(k,\dw\dw)
                  \end{pmatrix}\nonumber\\[8pt]
 \label{eq:psi_trafo}
                =&\,\hat{U}(k)\begin{pmatrix}
                    \psi(k,-) & 0 \\
                    0 & \psi(k,+)
                  \end{pmatrix}\hat{U}^{T}(-k)\ ,
\end{align}
where $\hat{U}^{T}$ is the transpose of $\hat{U}$.
Note that $\psi(k,\hel)\propto\langle b_{k\hel}b_{-k\hel}\rangle$ denotes the superconducting order parameter in helicity basis, while $\psi(k,ss')\propto\langle c_{ks}c_{-ks'}\rangle$ is the corresponding object in spin basis. The relation of the fermionic operators is given by
\begin{equation}
 \begin{pmatrix}
  c_{k\up}\\ c_{k\dw}
 \end{pmatrix}=\hat{U}(k)
 \begin{pmatrix}
  b_{k-}\\ b_{k+}
 \end{pmatrix}.
\end{equation}
Furthermore, we assumed here that we know $\psi(k,\hel)$ for any momentum $k$, while in reality we only have access to the projection onto the Fermi surface, $\psi(\hat{k},\hel)$. This issue will be addressed later.
From Eq.\,\eqref{eq:psi_trafo}, the particular elements of $\hat{\Delta}(k)$ are given by
\begin{align}
\psi(k,ss')=&\sum_{\hel}v_{s}(k,\hel)v_{s'}(-k,\hel)\psi(k,\hel)\nonumber\\
          =&\sum_{\hel}v_{s}(k,\hel)v_{s'}(-k,\hel)t(k,\hel)\psi^{(T)}(k,\hel)\\
    =&\sum_{\hel}\frac{1}{2}\bigg(\Big[\sigma_{0}+\hel\hat{\gamma}(k)\cdot\vec{\sigma}\Big]i\sigma_{y}\bigg)_{ss'}\psi^{(T)}(k,\hel)\nonumber \,
\end{align}
where $\hat{\gamma}=\vec{\gamma}/|\vec{\gamma}|$. The $d$-vector, which is defined by
\begin{equation}
 \hat{\Delta}(k)=(d_{\nu}\sigma^{\nu})i\sigma_{y}=
    \begin{pmatrix}
     -d_{x}+id_{y}  &   d_{0}+d_{z} \\
     -d_{0}+d_{z}   &   d_{x}+id_{y}
    \end{pmatrix},
\end{equation}
is then obtained as\,\cite{samokhin_gap_2008,sigrist_introduction_2009}
\begin{align}
\label{eq:d0}
 d_{0}(k)=&\,\frac{1}{2}\big[\psi^{(T)}(k,+)+\psi^{(T)}(k,-)\big]\ ,\\[5pt]
\label{eq:dv}
 \vec{d}(k)=&\,\frac{1}{2}\hat{\gamma}(k)\big[\psi^{(T)}(k,+)-\psi^{(T)}(k,-)\big]\ .
\end{align}
Eq.\,\eqref{eq:dv} reflects the result that $\vec{d}$ tends to be parallel to $\hat{\gamma}$, as long as the band splitting due to the Rashba SOC is larger than the superconducting gap, or in other words, $\alpha_{R}t_{1}>k_{B}T_{c}$\,\cite{frigeri_spin_2004,frigeri_superconductivity_2004,yokoyama-07prb172511,sigrist_introduction_2009}. Since the WCRG works at infinitesimal coupling, and in turn the SC gap is infinitesimally small as well, every finite value for $\alpha_R$ will result in $\vec{d}\parallel\hat{\gamma}$. With other words, the regime where $\alpha_{R}t_{1}<k_{B}T_{c}$ is not accessible within WCRG. This arises from the fact that we only consider Cooper pairs formed by electrons with opposite momentum, which must, hence, be from the same band, \ie contributions like $\psi(k,+,-)\propto\langle b_{k+}b_{-k-}\rangle$ are strictly zero.

It is important to note here that because inversion symmetry is broken by the Rashba SOC, the superconducting state can be a mix of singlet and triplet states\,\cite{gorkov_superconducting_2001,samokhin_gap_2008,sigrist_introduction_2009}. Furthermore, the order parameter matrix in spin space is not unitary anymore\,\cite{sigrist_introduction_2009}, \ie
\begin{align}
\label{eq:psi_psi_dagger}
 \hat{\Delta}\hat{\Delta}^{\dagger}=d_{\nu}d^{\nu*}\sigma_{0}+(d_{0}\vec{d}^{*}+d_{0}^{*}\vec{d})\cdot\vec{\sigma}\ .
\end{align}

For the specific system studied here, the Rashba SOC term, or more precisly the relation $\vec{d}||\hat{\gamma}$, enforces a specific structure on the $d$-vector. In particular, it yields $d_{z}=0$, since $\gamma_{z}=0$, and consequently there are no triplet states with zero total spin.
Before we apply this theory to the model Eq.\,\eqref{eq:total_general_ham} and present the results, we will discuss an example to demonstrate the implications from the above equations.
Let us consider the simplest case of a pure triplet state, which is given by $\psi^{(T)}(k,-)=-\psi^{(T)}(k,+)=\psi_{0}$. It becomes immediately apparent from Eq.\,\eqref{eq:d0} that $d_{0}=0$, \ie we have indeed a pure triplet state. Now, if we consider the square lattice with nearest neighbor hopping and Rashba SOC, we have
\begin{align}
 \gamma_{0}(k)=&-2t_{1}[\cos(k_{x})+\cos(k_{y})]\ ,\\[5pt]
 \gamma_{x}(k)=&+2t_{1}\alpha_{R}\sin(k_{y})\ ,\\[5pt]
 \gamma_{y}(k)=&-2t_{1}\alpha_{R}\sin(k_{x})\ ,
\end{align}
and thus
\begin{align}
 \psi(k,\up\up)=&+\psi_{0}\frac{\sin(k_{y})+i\sin(k_{x})}{\sqrt{\sin^{2}(k_{x})+\sin^{2}(k_{y})}}\ ,\\[5pt]
 \psi(k,\dw\dw)=&-\psi_{0}\frac{\sin(k_{y})-i\sin(k_{x})}{\sqrt{\sin^{2}(k_{x})+\sin^{2}(k_{y})}}\ .
\end{align}
This is, up to phase factors, a $p+ip$-wave for Cooper pairs with spin $1$ and a $p-ip$-wave for those with spin $-1$. That is, the superconducting quasi particles with spin up have an edge mode that propagates in one direction, whereas the ones with spin down propagate in the opposite direction, and thus preserve time-reversal symmetry. We can conclude from this consideration that the Rashba SOC term enforces the appearing triplet states to be helical, and forbids chiral states to appear.

\begin{table}[b]
 \caption{\label{tab:waves}List of low order lattice harmonics for each irreducible representation of the point group $D_{4}$.}
 \begin{tabular}{c|c|c}
  \hline
  irrep\ \  & label & base function \\
  \hline
  $A_{1}$ & $s$ & $1$ \\
  $A_{1}$ & ext. $s$ & $\cos(k_{x})+\cos(k_{y})$ \\
  $A_{2}$ & $g$ & $\sin(k_{x})\sin(k_{y})(\cos(k_{x})-\cos(k_{y}))$ \\
  $B_{1}$ & \ $d \equiv d_{x^{2}-y^{2}}$ & $\cos(k_{x})-\cos(k_{y})$ \\
  $B_{2}$ & $d' \equiv d_{xy}$ & $\sin(k_{x})\sin(k_{y})$ \\
  $E$ & $p$ & $\{\sin(k_{x}),\sin(k_{y})\}$ \\
  $E$ & ext. $p$ &\  $\{\sin(k_{x}),\sin(k_{y})\}(\cos(k_{x})+\cos(k_{y}))$ \\
  \multirow{2}{*}{$E$} & \multirow{2}{*}{$f$} & $\{\sin(k_{x})(\cos(k_{x})-\cos(k_{y})),$ \\
  & & \phantom{\{}$\sin(k_{y})(\cos(k_{x})-\cos(k_{y}))\}$\\
  \hline
 \end{tabular}
\end{table}
More generally, we can see which mixed states can be expected to appear. Eqs.\,\eqref{eq:gamma_TR_phase} and \eqref{eq:gamma_TR_phase_odd} state that the vertex function using time reversal partners has to be even in momentum, which means that its eigenfunctions, $\psi^{(T)}(k,\hel)$, have to transform according to an even irreducible representation (irrep) of the symmetry group of the lattice. Hence, on the square lattice, available irreps are $A_{1}$, $A_{2}$, $B_{1}$, and $B_{2}$, which correspond in terms of lowest order lattice harmonics to (extended) $s$-, $g$-, $d$-, and $d'$-wave, respectively. Multiplying $\vec{\gamma}$ to those functions to obtain the $d$-vector, as done above, leads to the following mixed singlet-triplet states:
\begin{equation}\label{eq:mixed-states}
\begin{aligned}
 A_{1}:&\,\,s+p\\
 A_{2}:&\,\,g+f\\
 B_{1}:&\,\,d+f\\
 B_{2}:&\,\,d'+p
\end{aligned}
\end{equation}
where the first term denotes the symmetry of the singlet component ($s$, $d$, and $g$) and the second one the helical triplet component ($p$ and $f$). The latter can be explicitly expressed as
\begin{equation}
 p:\hspace{4mm}\psi(k,\up\up)\,\hat{=}\,p+ip\ ,\hspace{4mm}\psi(k,\dw\dw)\,\hat{=}\,p-ip\ ,
\end{equation}
and similar for the $f$-wave. Note that $p$-wave might correspond both to standard and extended $p$-wave.
A list of base functions for some low order harmonics is given in Tab.\,\ref{tab:waves}.

Finally, we are left with a technical problem: we know the form factor, $\psi(k,\hel)$, only for momenta on the Fermi surface, \ie $\psi(\hat{k},\hel)$. However, to calculate the $d$-vector for any given momentum $k$, Eqs.\,\eqref{eq:d0} and\ \eqref{eq:dv} demands the formfactor of both helicities, \ie bands, at the \emph{same} momentum $k$, one of which cannot be on the Fermi surface. To solve this problem, we extend the form factor to the whole Brillouin zone by fitting a set of lattice harmonics to the respective Fermi surfaces (details on the fitting process are given in App.\,\ref{sec:appA}).

%%%%%%%%%%%%%%%%%%%%%%%%%%%%%%%%%%%%%%%%%%%%%%%%%%%%%%%%%%%%%%%%%
%
%                                 R E S U L T S
%
%%%%%%%%%%%%%%%%%%%%%%%%%%%%%%%%%%%%%%%%%%%%%%%%%%%%%%%%%%%%%%%%%
\section{Results}
\label{sec:results}
We study the square lattice with nearest and next nearest neighbor hopping ($t_{1}$ and $t_{2}$), repulsive onsite interaction ($U_{0}$), and Rashba SOC ($\alpha_{R}t_1$ and $\alpha_{R}t_2$). The bandstructure of the non-interacting system is thus given by
\begin{equation}
 E(k\hel)=\gamma_{0}+\hel\sqrt{\gamma_{x}^{2}+\gamma_{y}^{2}}
 \end{equation}
 with
 \begin{align}
 \gamma_{0}=&-2t_{1}\big[\cos(k_{x})+\cos(k_{y})\big]-4t_{2}\cos(k_{x})\cos(k_{y})\ ,\nonumber\\[5pt]
 \gamma_{x}=&+\alpha_{R}\big[2t_{1}\sin(k_{y})+4t_{2}\cos(k_{x})\sin(k_{y})\big]\ ,\nonumber\\[5pt]
 \gamma_{y}=&-\alpha_{R}\big[2t_{1}\sin(k_{x})+4t_{2}\sin(k_{x})\cos(k_{y})\big]\ .\nonumber
\end{align}
%
%%%%%%%%%%%%%%%%%%%%%%%%%%%%%%%%%%%%%%%%%%%%%%%%%%%%%%%%%%%%%%
%           F I G. 2
%%%%%%%%%%%%%%%%%%%%%%%%%%%%%%%%%%%%%%%%%%%%%%%%%%%%%%%%%%%%%%
\begin{figure}[t!]
 \includegraphics[width=0.98\columnwidth]{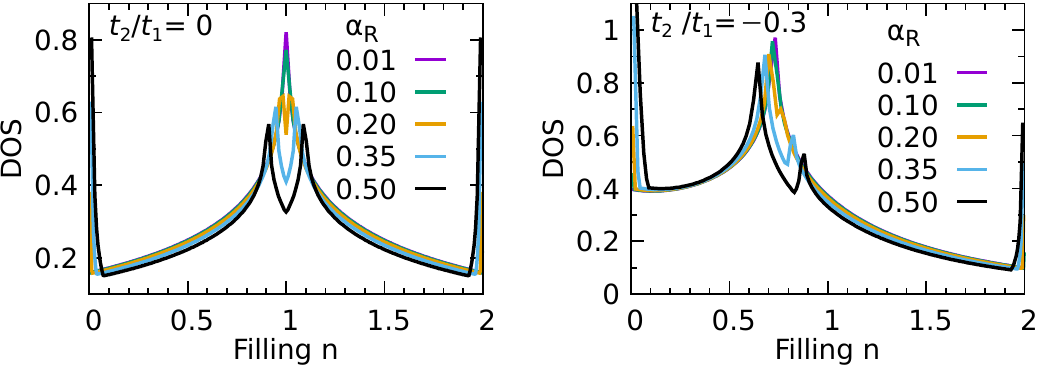}
 \caption{\label{fig:DOS}Density of states (DOS) of the square lattice with Rashba SOC as a function of filling $n$, for different strengths of the SOC, $\alpha_{R}$. The second neighbor hopping is given by $t_{2}/t_{1}=0$ (left) and $t_{2}/t_{1}=-0.3$ (right).}
\end{figure}
%%%%%%%%%%%%%%%%%%%%%%%%%%%%%%%%%%%%%%%%%%%%%%%%%%%%%%%%%%%%%%
%

The DOS, $\rho$, of the system for next-nearest neighbor hopping $t_{2}=0$ and $t_{2}=-0.3$ is shown in Fig.\,\ref{fig:DOS}. In both cases the van-Hove singularity splits into two, one for each helicity. The singularities appear at different values for the filling, $n_{{\rm vH},\hel}$, for non-zero SOC, given by
\begin{align}
\label{eq:nvH}
 n_{{\rm vH},\hel}=&\int_{-\infty}^{\mu_{{\rm vH},\hel}}\rho(\varepsilon){\rm d}\varepsilon\ ,\\
\label{eq:muvH}
 \mu_{{\rm vH},\hel}=&\,\hel\cdot2t_{1}\Big(1-\sqrt{1+\alpha_{R}^{2}}\Big)+4t_{2}\sqrt{1+\alpha_{R}^{2}}\ ,
\end{align}
for $\zeta = \pm 1$.
Note that Eq.\,\eqref{eq:muvH} is only valid for $|t_{2}|\leq0.5|t_{1}|$.
The difference between $n_{{\rm vH},+}$ and $n_{{\rm vH},-}$ increases with $\alpha_{R}$, as shown in Fig.\,\ref{fig:DOS}.

\begin{figure}[t!]
 \includegraphics[width=0.9\columnwidth]{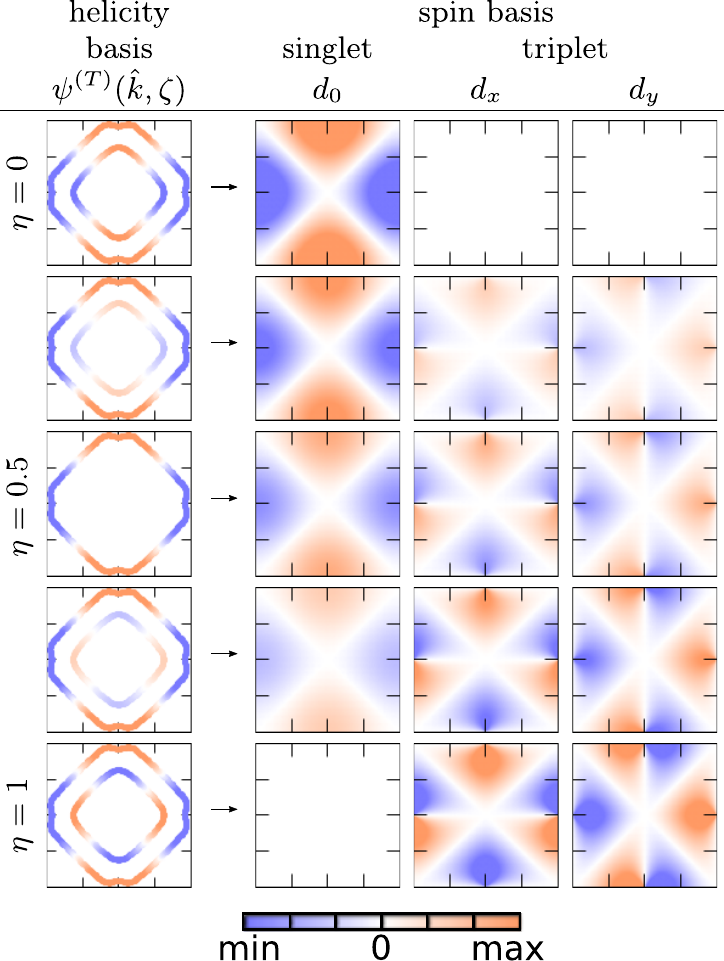}
 \caption{\label{fig:showcase}Transition from pure $d_{x^{2}-y^{2}}$-wave to pure helical $f$-wave. This is done by setting $\psi^{(T)}(k,\hel)=c_{\hel}[\cos(k_{x})-\cos(k_{y})]$ and varying from $c_{+}=c_{-}$ (top row) via $c_+=0$ (middle row) to $c_{+}=-c_{-}$ (bottom row).}
\end{figure}

We will discuss the leading superconducting instabilities obtained within the WCRG method as a function of band filling, $n$, and for different strengths of the Rashba SOC, $\alpha_{R}\in\{0,0.01,0.1,0.2,0.35,0.5\}$. 
The main difference between this analysis with and without Rashba SOC is that in the former case we find the leading instability in helicity space and need to transform this result back to spin space.
In the following  we will first give a simple example to demonstrate how the resulting form factor obtained in helicity basis, $\psi(k,\hel)$, relates to the $d$-vector in spin basis.
We assume that we have a pure $d_{x^{2}-y^{2}}$ as the form factor on both Fermi surfaces, \ie
\begin{align}
\label{eq:showcase_psi1}
 \psi^{(T)}(k,+)=&\,c_{+}[\cos(k_{x})-\cos(k_{y})]\ ,\\[5pt]
\label{eq:showcase_psi2}
 \psi^{(T)}(k,-)=&\,c_{-}[\cos(k_{x})-\cos(k_{y})]\ .
\end{align}
Note that both $\psi$'s are even functions in momentum. However, we still have the freedom to choose different values for the prefactors $c_{\pm}$, and this freedom is exactly what enables us to obtain triplet states, which are odd in momentum, as given by Eq.\,\eqref{eq:dv}. Varying $c_{+}$ smoothly from the value of $c_{-}$ to $-c_{-}$ thus yields a transition between a pure singlet state ($d_{x^{2}-y^{2}}$-wave) at $c_{+}=c_{-}$ to a pure triplet state (helical $f$-wave) at $c_{+}=-c_{-}$, which is shown in Fig.\,\ref{fig:showcase}. For $c_{+}\neq\pm c_{-}$, we have a mixed state with singlet \emph{and} triplet components.

%
%%%%%%%%%%%%%%%%%%%%%%%%%%%%%%%%%%%%%%%%%%%%%%%%%%%%%%%%%%%%%%
%           F I G. 3
%%%%%%%%%%%%%%%%%%%%%%%%%%%%%%%%%%%%%%%%%%%%%%%%%%%%%%%%%%%%%%
%%%%%%%%%%%%%%%%%%%%%%%%%%%%%%%%%%%%%%%%%%%%%%%%%%%%%%%%%%%%%%
%

\begin{figure}[b!]
 \includegraphics[width=0.65\columnwidth]{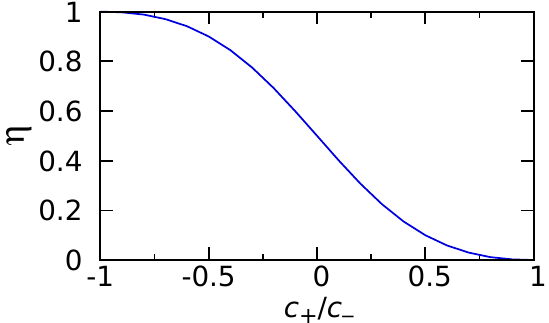}
 \caption{\label{fig:showcase_triplet}Triplet contribution, $\eta$, to the state described in Eqs.\,\eqref{eq:showcase_psi1} and \eqref{eq:showcase_psi2} as a function of $c_{+}/c_{-}$.}
\end{figure}

We evaluate the triplet contribution, $\eta$, to the full state by calculating the average over the Brillouin zone
\begin{align}
\label{eq:theta_tot}
 \eta&\equiv\,\frac{\int_{\rm BZ}\frac{{\rm d}^{2}k}{(2\pi)^{2}}\Big(|d_{x}(k)|^{2}+|d_{y}(k)|^{2}\Big)}{\int_{\rm BZ}\frac{{\rm d}^{2}k}{(2\pi)^{2}}\Big(|d_{0}(k)|^{2}+|d_{x}(k)|^{2}+|d_{y}(k)|^{2}\Big)\ }\ .
\end{align}
Note that the integrals run over the entire Brillouin zone; after the necessary fitting process (see above and App.\,\ref{sec:appA}), the $\vec d$ vector is given in the whole Brillouin zone instead of just on the Fermi surface.
By construction, $0\leq \eta \leq 1$.
The triplet contribution to the state described in Eqs.\,\eqref{eq:showcase_psi1} and \eqref{eq:showcase_psi2} as a function of $c_{+}/c_{-}$ is shown in Fig.\,\ref{fig:showcase_triplet}.
%
%%%%%%%%%%%%%%%%%%%%%%%%%%%%%%%%%%%%%%%%%%%%%%%%%%%%%%%%%%%%%%
%           F I G. 4
%%%%%%%%%%%%%%%%%%%%%%%%%%%%%%%%%%%%%%%%%%%%%%%%%%%%%%%%%%%%%%
%%%%%%%%%%%%%%%%%%%%%%%%%%%%%%%%%%%%%%%%%%%%%%%%%%%%%%%%%%%%%%
%
%
%%%%%%%%%%%%%%%%%%%%%%%%%%%%%%%%%%%%%%%%%%%%%%%%%%%%%%%%%%%%%%
%           F I G. 5
%%%%%%%%%%%%%%%%%%%%%%%%%%%%%%%%%%%%%%%%%%%%%%%%%%%%%%%%%%%%%%
\begin{figure*}[t!]
 \includegraphics[width=2.04\columnwidth]{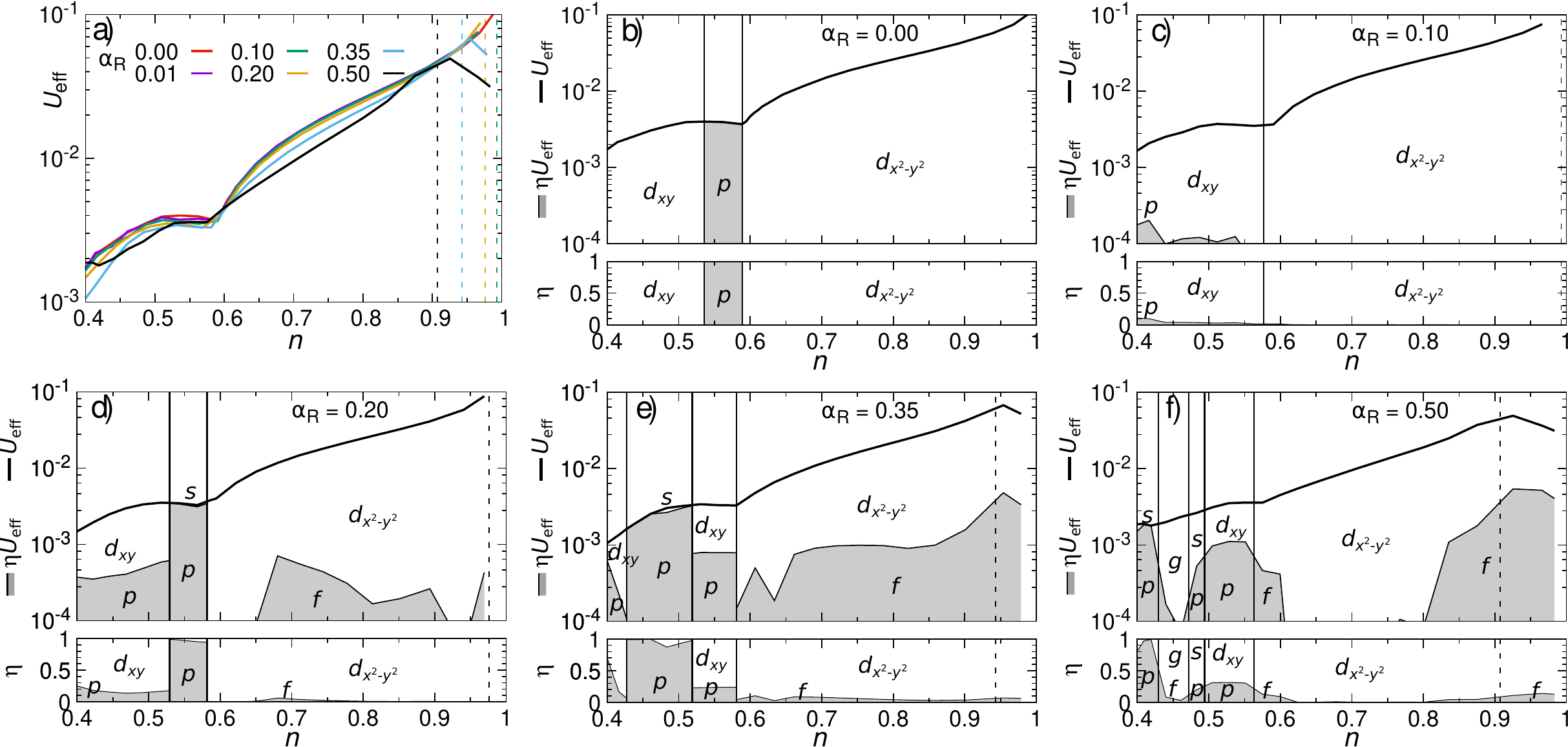}
 \caption{\label{fig:Uefft_2=0}
(a) $U_{\rm eff}$ as function of band filling, $n$, for different SOC strengths, $\alpha_{R}$. 
(b)-(f) Superconducting phase diagrams for selected values of $\alpha_R$. In addition to $U_{\rm eff}$ vs.\ $n$, the symmetry of the leading instability is shown.
The shaded region indicates the triplet contribution, $\eta\,U_{\rm eff}$, to the full order parameter, whereas the pure ratio $\eta$ is shown below the panels. The van-Hove singularities are indicated by dashed lines. 
 %For non-zero SOC the $d$-vector is shown in the form of contour plots for three values of $n$, as indicated by the arrows. 
% The respective order parameters for zero SOC ($\alpha_{R}=0$) are, from left to right: $d_{xy}$, $p_{x}+ip_{y}$, $d_{x^{2}-y^{2}}$\,\cite{wolf_unconventional_2018}.
 }
\end{figure*}
%%%%%%%%%%%%%%%%%%%%%%%%%%%%%%%%%%%%%%%%%%%%%%%%%%%%%%%%%%%%%%
%

If superconductivity only occurs on one of the two Fermi surfaces, \ie $c_{+}=0$ or $c_{-}=0$, we obtain a state which is half singlet and half triplet ($\eta=0.5$) as shown in the middle row of Fig.\,\ref{fig:showcase}.

It is important to note here that $\eta$ just serves as an estimation for the ratio of singlet and triplet contributions to the full superconducting order parameter, since the denominator in Eq.\,\eqref{eq:theta_tot} does \emph{not} correspond to the full contribution of $\hat{\Delta}\hat{\Delta}^{\dagger}$ given in Eq.\,\eqref{eq:psi_psi_dagger}; it is, however, a good approximation.

We further note that the results from the WCRG
often require some higher harmonics to provide a good fit to the superconducting form factor instead of a single lattice harmonic. Higher lattice harmonics can play a non-trivial role. Usually, this does not change the physics qualitatively, but the technical details become more involved (see App.\,\ref{sec:appA}).

Now we have all the tools at hand to discuss the results obtained by solving Hamiltonian\,\eqref{eq:total_general_ham}. The resulting effective interactions, $U_{\rm eff}$, for $t_{2}=0$ and as a function of filling $n$ are shown in Fig.\,\ref{fig:Uefft_2=0} for several relevant values in the range $0 \leq \alpha_{R} \leq 0.5$. 
Firstly, we see that the overall effective interaction does not change much with increasing spin-orbit coupling [panel\,(a)]. Two notable exceptions can be observed: 
1) above the van-Hove singularity, which moves to lower fillings as $\alpha_{R}$ increases (\cf Eqs.\,\eqref{eq:nvH}, \eqref{eq:muvH} and Fig.\,\ref{fig:DOS}), one observes a decrease of $U_{\rm eff}$ for increasing SOC. This decrease stems from the accompanied sharp decrease of the DOS as a function of $\alpha_{R}$ in this region. 
2) for fillings in the range $n\in[0.62,0.85]$ one also observes a slight decrease of $U_{\rm eff}$ with increasing $\alpha_{R}$.
In panels (b)--(f) of Fig.\,\ref{fig:Uefft_2=0} we show the phase diagrams for individual values of $\alpha_R$. In particular, panel\,(b) corresponds to the spin-orbit-free case, $\alpha_R=0$ [see Refs.\,\onlinecite{wolf_unconventional_2018,raghu_superconductivity_2010}]. The singlet part of the superconducting condensate is shown in white while the triplet part, $\eta U_{\rm eff}$ defined in Eq.\eqref{eq:theta_tot}, is shown as a shaded area. Moreover, we show the mixing ratio of singlet and triplet states, $\eta$, separately in the small panels below the main panels (b)--(f). We note that mixing is forbidden for $\alpha_{R}=0$ (panel\,(b)), and here $\eta$ can only take the values 0 (singlet) or 1 (triplet). The phase diagram for $\alpha_R=0$ contains spin-singlet $d_{xy}$-wave order for small $n$. Then there is a small intermediate spin-triplet phase for $0.54\leq n\leq 0.59$ with $p_x + i p_y$-wave order, followed by another spin-singlet phase with $d_{x^2-y^2}$-wave order  up to van Hove filling $n=1$.

Turning on Rashba SOC $\alpha_R$ allows  for mixing of singlet and triplet phases. Panel (c) of Fig.\,\ref{fig:Uefft_2=0}  shows results for $\alpha_R=0.1$ (being representative for the entire range $0<\alpha_R \leq 0.1$) where only small tripet contributions with $p$-wave symmetry are mixed into the $d_{xy}$-wave order for small $n$. The chiral $p$-wave phase for $0.54\leq n\leq 0.59$ present at $\alpha_R=0$ is absent, the reason for which we will explain below in detail. Thus we are left with two phases with dominant but different $d$-wave symmetries.
In panel (d) we show $\alpha_R=0.2$: while the original structure of the $\alpha_R=0$ phase diagram is still visible, now significant triplet contributions are mixed into the singlet phases and small singlet contributions into the triplet phase. We note that the mixing ratio $\eta$, shown below the main panel, takes arbitrary values between 0 and 1, as expected. The mixed condensates realized for $\alpha_R=0.2$ correspond to $d'+p$, $s+p$ and $d+f$ (\cf Eq.\,\eqref{eq:mixed-states}). Note that the $s+p$ phase, which corresponds to the chiral $p+ip$ phase for $\alpha_{R}=0$, is not only mixed with a singlet $s$-wave, but also the triplet part changes to a helical $p$-wave, for the reasons outlined above.
Panel (e) shows $\alpha_R=0.35$ where the mixing of singlet and triplet phases dominates but also the deformation of the FSs due to Rashba SOC has altered the structure of the $\alpha_R=0$ phase diagram at small $n$.
Finally, $\alpha_R=0.5$ is shown in Fig.\,\ref{fig:Uefft_2=0}\,(f): here we find many small phases with different singlet-triplet mixtures. In contrast to smaller values of $\alpha_R$, here we also detect $g$-wave order with small $f$-wave contributions.

%%%%%%%%%%%%%%%%%%%%%%%%%%%%%%%%%%%%%%%%%%%%%%%%%%%%%%%%%%%%%%
%           F I G. 6
%%%%%%%%%%%%%%%%%%%%%%%%%%%%%%%%%%%%%%%%%%%%%%%%%%%%%%%%%%%%%%
\begin{figure}[t!]
 \includegraphics[width=0.99\columnwidth]{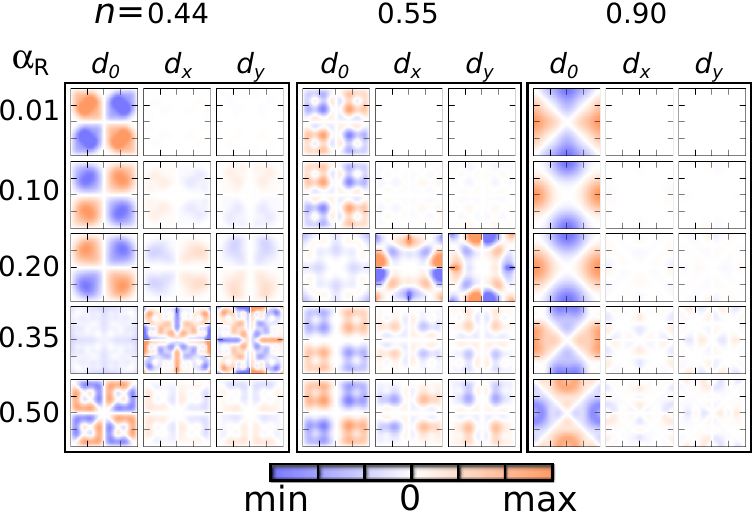} %ratio_eval_t2=0_all_v3.pdf}
 \caption{\label{fig:Ueff_formfactors}
For non-zero SOC the $d$-vector is shown in the form of contour plots in the Brillouin zone for three relevant values of $n$ and selected values of $\alpha_R$.}
\end{figure}
%%%%%%%%%%%%%%%%%%%%%%%%%%%%%%%%%%%%%%%%%%%%%%%%%%%%%%%%%%%%%%

Furthermore, we can see from Fig.\,\ref{fig:Uefft_2=0} %and\,\ref{fig:Ueff_L_SL} 
that for extended regions the ratio $\eta$ tends to stay away from $0$ or $1$ with increasing $\alpha_{R}$, \ie the mixing of singlet and triplet states becomes more pronounced for stronger spin-orbit coupling. In the region $n\approx[0.65,0.83]$, however, $\eta$ increases at first, until $\alpha_{R}=0.35$, and then decreases again for $\alpha_{R}=0.5$. 
%While $\eta$ tends to be non-zero for non-zero $\alpha_R$, it does not increase monotonically with $\alpha_R$
Hence, there is an optimal value for $\alpha_{R}$ for maximising the triplet contribution to the dominant singlet state in this region.

%%%%%%%%%%%%%%%%%%%%%%%%%%%%%%%%%%%%%%%%%%%%%%%%%%%%%%%%%%%%%%
%           F I G. 7
%%%%%%%%%%%%%%%%%%%%%%%%%%%%%%%%%%%%%%%%%%%%%%%%%%%%%%%%%%%%%%
\begin{figure}[t]
 \includegraphics[width=0.85\columnwidth]{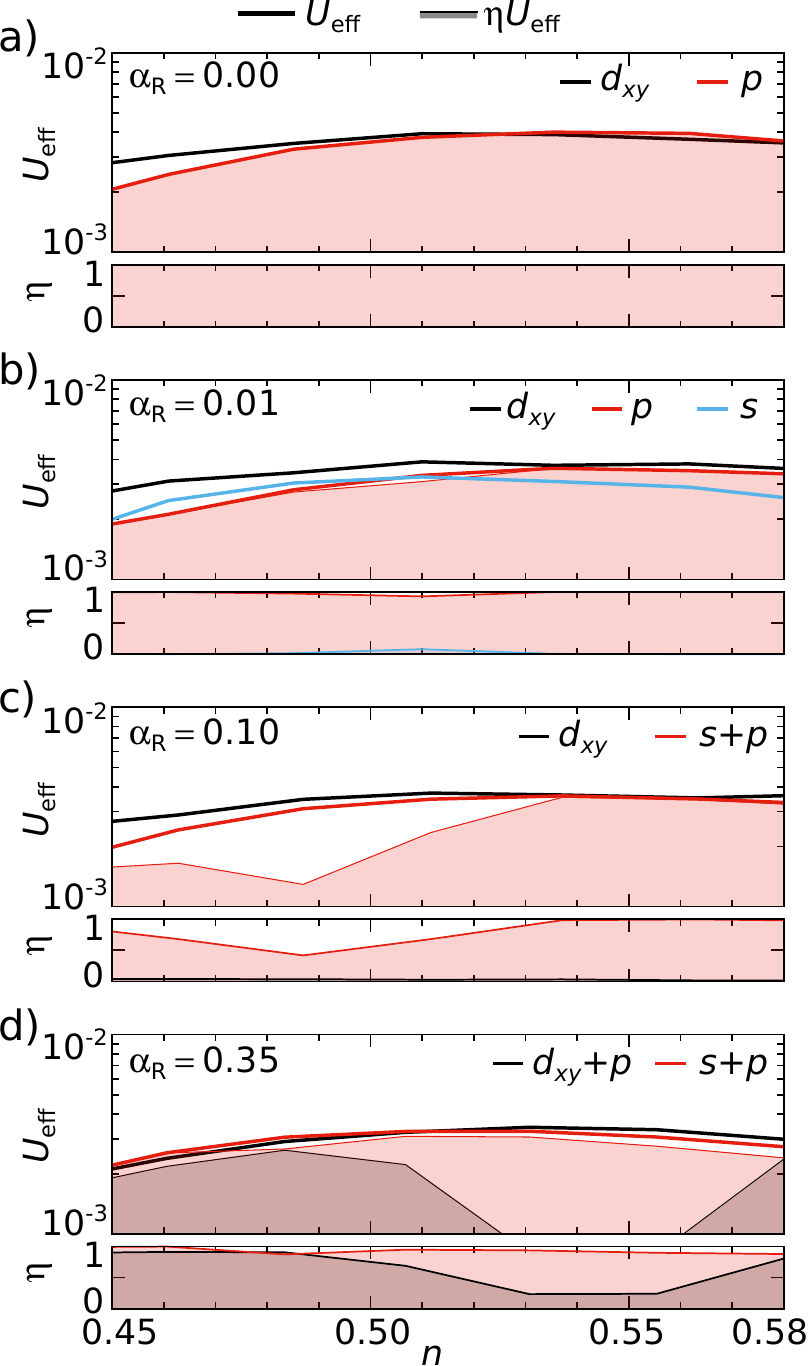}
 \caption{\label{fig:Ueff_L_SL}$U_{\rm eff}$ as function of band filling, $n$, for different SOC-strength, $\alpha_{R}$, in the region $n\in [0.45,0.58]$, where the $d$- and $p$-wave dominant solutions are almost degenerate. Partially, also an extended $s$-wave singlet state is competing with the former two. In the presence of SOC, $d$ ($p$) becomes a $d+p$ ($s+p$) mixture.
 The shaded region (both in red and black) indicates the triplet contribution, $\eta\,U_{\rm eff}$, to the full order parameter. It is apparent that the almost-degeneracy between the $d$- and $p$- waves in this region persists even for strong spin-orbit coupling.}
\end{figure}
%%%%%%%%%%%%%%%%%%%%%%%%%%%%%%%%%%%%%%%%%%%%%%%%%%%%%%%%%%%%%%

In addition to $U_{\rm eff}$ and the singlet triplet mixture in Fig.\,\ref{fig:Uefft_2=0}, we show in Fig.\,\ref{fig:Ueff_formfactors} representative $d$-vectors as contour plots corresponding to the three different fillings $n=0.44, 0.55, 0.9$ and to different values of $\alpha_{R}$. The contour plots reveal the symmetry of the irreps, associated with the leading instability, in the Brillouin zone.

As mentioned before, for $\alpha_{R}=0$ the mixing of singlet and triplet states is forbidden, \ie $\eta\equiv 0$ (singlet) or $\eta\equiv 1$ (triplet). As we can see in Fig.\,\ref{fig:Uefft_2=0}\,(c), $\eta U_{\rm eff}$ tends to zero everywhere in the limit $\alpha_{R}\rightarrow0$, \ie to the spin-orbit-free case. An exception is observed in the region where the pure triplet state is found in the absence of SOC ($n\approx[0.54,0.59]$, indicated by the shaded area in Fig.\,\ref{fig:Uefft_2=0}\,(b)).

Here $\eta U_{\rm eff}$ is not approaching $U_{\rm eff}$ for small values of $\alpha_R$. Fig.\,\ref{fig:Uefft_2=0}\,(c), but also other results in the regime $0<\alpha_R\leq 0.1$ (not shown here) show the suppression of the triplet $p$-wave phase. This discrepancy has a simple explanation: in the mentioned regime we find a (near-) degeneracy of two or even three superconducting instabilities. Minimal deformations of the FS, as caused by small changes of $\alpha_R$, let these almost degenerate instabilities change their ordering. In a previous work, we have shown that even the numerical resolution for evaluating the integrals affects which of the nearly-degenerate instabilities wins [see Sec.\,IV in Ref.\,\onlinecite{wolf_unconventional_2018}]. In Fig.\,\ref{fig:Ueff_L_SL} we plot the leading and subleading instabilities for different values of $\alpha_{R}$ in the region $n=[0.45,0.58]$. This almost-degeneracy persists even for larger values of $\alpha_{R}$, and we find that in the regime $0.01 \leq \alpha_R \leq 0.15$ the triplet state swaps position with another state which is of pure singlet type. The solid lines indicating $U_{\rm eff}$ of the competing pairing channels are extremely close together (note the logarithmic scale), which causes the abrupt changes from singlet to triplet states or vice versa. For instance, for $\alpha_R=0$ [panel\,(a)] there is a pure singlet state with $d_{xy}$-wave order (shown in black) competing with a pure triplet state with $p$-wave order (shown in red), leading to the phase diagram Fig.\,\ref{fig:Uefft_2=0}\,(b). For $\alpha_R=0.01$, there are even three competing instabilities, two pure singlet states with $d_{xy}$-wave and extended $s$-wave order and the previously mentioned triplet state with $p$-wave order. While the $p$-wave state wins for $\alpha_R=0$, it looses by a hair-split for $\alpha_R=0.01$ and $0.1$ [\cf Fig.\,\ref{fig:Uefft_2=0}\,(c)].  
%Similar behavior can be observed up to quite large SOC $\alpha_R=0.35$ where an almost perfect degeneracy can be observed.
Similarly, up to quite large SOC $\alpha_R=0.35$ one can observe this almost perfect degeneracy.

Three comments are in order: (i) this issue of nearly-degenerate states and competing pairing channels is not specific to the presence of Rashba SOC, but a rather generic problem. (ii) in fact, ``almost-degeneracies'' are not a {\it problem} but rather a feature: if such degeneracies are sufficiently robust, they can lead to two-component order parameters. For instance, when a $d_{xy}$-wave and another $d_{x^2-y^2}$-wave state on the square lattice become (almost) degenerate, they can form complex superpositions of the type $d+id$, resulting in chiral, topological superconductivity. In the present example, even a highly exotic three-component order parameter of the type $d_{xy} + \alpha p_x + \beta p_y$ is possible (with $\alpha$, $\beta$ complex constants). (iii) we emphasize that these results were obtained in the weak coupling regime; whether or not such near-degeneracies are stable upon increasing Coulomb interactions is less clear and requires further investigations on a case-by-case basis using other methods.

In summary, the investigation of the nearest-neighbor square lattice Hubbard model in the presence of Rashba SOC -- being the most paradigmatic lattice model to study -- reveals already a rich phenomenology. We note that the overall amplitude $U_{\rm eff}$ and thus the critical transition temperature $T_c$ hardly changes with varying $\alpha_R$. Spin-singlet superconducting phases, which dominate the phase diagram in the absence of SOC, mix with spin-triplet states as $\alpha_R$ increases. We emphasize that the increase of triplet contribution is non-monotonic in $\alpha_R$ and an optimal value for maximizing triplet contributions exists.

%%%%%%%%%%%%%%%%%%%%%%%%%%%%%%%%%%%%%%%%%%%%%%%%%%%%%%%%%%%%%%%%%
%
%                                D I S C U S S I O N
%
%%%%%%%%%%%%%%%%%%%%%%%%%%%%%%%%%%%%%%%%%%%%%%%%%%%%%%%%%%%%%%%%%

\section{Discussion}
\label{sec:discussion}

In this section, we discuss the following five aspects of our work: (i) we benchmark to results in the literature; (ii) we briefly discuss the experimental relevance; (iii) we address the interesting topic of topological phase transitions; (iv) we give an outlook what to expect when considering other lattices; (v) we briefly discuss how random disorder might affect our results.

%%%%%%%%%%%%%%%%%%%%%%%%%%%%%%%%%%%%%%%%%%%%%%%%%%%%%%%%%%%%%%
%           F I G. 8
%%%%%%%%%%%%%%%%%%%%%%%%%%%%%%%%%%%%%%%%%%%%%%%%%%%%%%%%%%%%%%
\begin{figure}[t]
 \includegraphics[width=0.9\columnwidth]{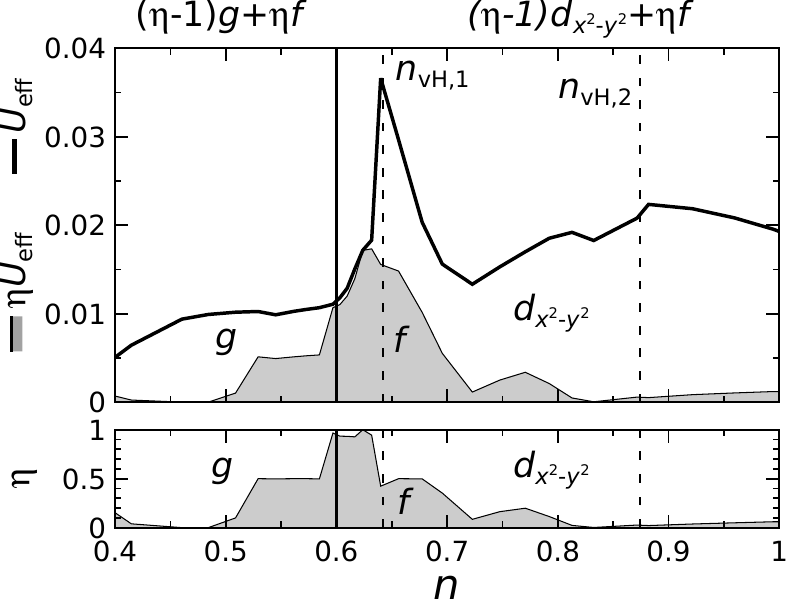}
 \caption{\label{fig:Ueff_t2_03}$U_{\rm eff}$ as function of band filling, $n$, for $t_{2}/t_{1}=-0.3$ and $\alpha_{R}=0.5$, in the region $n\in [0.4,1]$. The shaded region indicates the triplet contribution, $\eta\,U_{\rm eff}$, to the full order parameter. In the lower panel we plot $\eta$ as a function of filling $n$. The van Hove singularities, $n_{\rm vH}$, are indicated as dashed lines. For $n<0.6$ we find a mix of $g$- and helical $f$-wave, for fillings $0.6<n<0.624$ almost pure helical $f$-wave, and for $n>0.624$ a mix of $d_{x^{2}-y^{2}}$ and helical $f$-wave.}
\end{figure}
%%%%%%%%%%%%%%%%%%%%%%%%%%%%%%%%%%%%%%%%%%%%%%%%%%%%%%%%%%%%%%

Let us consider non-zero $t_{2}$ in order to study the effect of longer ranged hopping on the mixed singlet-triplet states. For the sake of benchmarking and comparing with results in the literature\,\cite{greco_mechanism_2018}, we choose the bandstructure parameters to be
\begin{equation}\label{CeCoIn}
 t_{2}/t_{1}=-0.3,\hspace{4mm} \alpha_{R}=0.5\ .
\end{equation}
In Ref.\,\onlinecite{greco_mechanism_2018} the very same model was investigated by virtue of random phase approximation (RPA). Note that in Ref.\,\onlinecite{greco_mechanism_2018} a factor 4 is missing in the second neighbor hopping term, $t'\equiv t_{2}$; this can be confirmed by comparing DOS plots.
According to Eq.\,\eqref{eq:nvH}, the van Hove singularities are located at $n_{\rm vH,1}\approx0.642$ and $n_{\rm vH,2}\approx0.874$. The results for $U_{\rm eff}$ and $\eta$ are shown in Fig.\,\ref{fig:Ueff_t2_03}.

\begin{figure}[t!]
 \includegraphics[width=0.9\columnwidth]{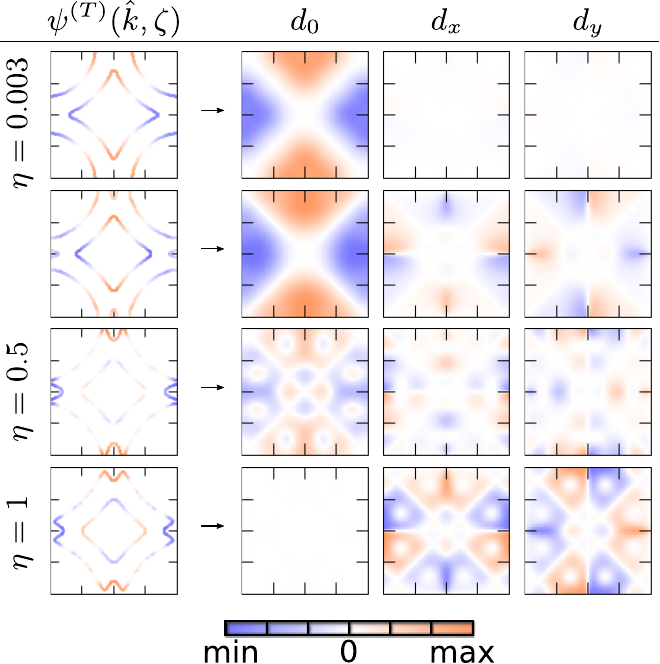}
 \caption{\label{fig:showcase_real}
Transition from pure $d_{x^{2}-y^{2}}$-wave to pure helical $f$-wave as a function of the filling $n$. These are results for $t_{2}/t_{1}=-0.3$, $\alpha_{R}=0.5$, and, from top to bottom, $n=\{0.832,\,0.748,\,0.656,\,0.624\}$ and $\eta=\{0.003,\,0.164,\,0.5,\,1\}$.}
\end{figure}

In agreement with Greco and Schnyder\,\cite{greco_mechanism_2018}, we find a strong peak of $f$-wave around the first van Hove singularity, $n_{\rm vH,1}$. For lower fillings, until $n=0.6$, we find an almost pure helical $f$-wave state, which gets than gradually mixed with $g$-wave (not considered in Ref.\,\onlinecite{greco_mechanism_2018}) for decreasing filling, where the $f$-wave contribution becomes negligible for $n<0.5$. Above the van Hove singularity, $n>n_{\rm vH,1}$, the $f$-wave gets gradually mixed with $d_{x^{2}-y^{2}}$-wave, starting from equal contributions from both, $f$- and $d$-wave right above $n_{\rm vH,1}$ to almost pure $d_{x^{2}-y^{2}}$-wave at $n\approx0.832$. Starting from this point of a pure singlet state, decreasing the filling lets us thus replicate the theoretical study in Fig.\,\ref{fig:showcase}, \ie we can adjust the mixing ratio $\eta$ gradually by changing the filling $n$, which is shown in Fig.\,\ref{fig:showcase_real}.
The regime below $n=1$ is also in Ref.\,\onlinecite{greco_mechanism_2018} dominated by $d_{x^2-y^2}$-wave order; however, below $n_{\rm vH,2}$ the dominating pairing channel reported in Ref.\,\onlinecite{greco_mechanism_2018} is a $d_{xy}$-wave order. This disagreement might stem from the different interaction strengths and the methodological differences in both works.

Most recently, the Rashba-Hubbard model with parameters \eqref{CeCoIn} was also studied in Ref.\,\onlinecite{ghadimi-19prb115122} using RPA. However, they considered a slightly different bandstructure, in particular, without the factor 4 in the second neighbor hopping term $t_{2}$. Thus, their results correspond to $t_{2}/t_{1}=-0.075$ in our notation. Given the proximity to $t_{2}=0$, we can at least qualitatively compare their results to Fig.\,\ref{fig:Uefft_2=0}\,(f). We both find a dominant $d_{x^{2}-y^{2}}$ phase in the same doping regime. For smaller doping they also find a $g$-wave phase. More interestingly, Ref.\,\onlinecite{ghadimi-19prb115122} also calculated a singlet-triplet ratio, similar to $\eta$. Around $n=0.4$, they find a dominant triplet phase, which matches our results. Around $n=0.6$, they find an almost purely singlet state, which we have at around $n=0.65$. Given the small difference in $t_{2}$ between the two works, the agreement is satisfying.

%
%%%%%%%%%%%%%%%%%%%%%%%%%%%%%%%%%%%%%%%%%%%%%%%%%%%%%%%%%%%%%%
%           F I G. 9
%%%%%%%%%%%%%%%%%%%%%%%%%%%%%%%%%%%%%%%%%%%%%%%%%%%%%%%%%%%%%%
%%%%%%%%%%%%%%%%%%%%%%%%%%%%%%%%%%%%%%%%%%%%%%%%%%%%%%%%%%%%%%
%

Often unconventional superconductors contain heavy elements which is favorable for non-negligible spin-orbit interactions. For instance, Sr$_2$RuO$_4$\,\cite{mackenzie-03rmp657}, Cu$_x$Bi$_2$Se$_3$\,\cite{hor-10prl057001,sasaki-11prl217001} and CePt$_3$Si\,\cite{bauer-04prl027003,yanase-08jpsp124711,smidman-17rpp036501} are the standard candidate materials for topological superconductivity. In particular, Sr$_2$RuO$_4$ was believed to realize spin-triplet $p$-wave pairing. Due to recent experimental progress\,\cite{pustogow-19n72,ghosh_natphys_2020} it became apparent that a two-component spin-singlet order parameter represents a more likely scenario\,\cite{kivelson-20arXiv,suh_stabilizing_2019}. One of the recent theory proposals\,\cite{suh_stabilizing_2019} is based on the presence of significant Rashba SOC.

Superlattices with tunable layer thickness provides another example of unconventional superconductivity where Rashba SOC is important. The heavy-fermion superlattice CeCoIn$_5$/YbCoIn$_5$\,\cite{mizukami-11np849,shimozawa-14prl156404} is such a candidate where inversion symmetry is broken by the modulation of the layer thickness of the superlattice. In fact, the parameters discussed above, Eq.\,\eqref{CeCoIn} and Ref.\,\onlinecite{greco_mechanism_2018}, are motivated by this compound. We would like to emphasize, however, that the aim of this work is not to model a specific material but rather to present a principal, thorough study of how Rashba SOC affects the superconducting instabilities; therefore we have chosen the simplest and most paradigmatic model one can think of.

The mixing of singlet and triplet phases leads to an interesting point of study: suppose that the singlet phase is a gapped phase with zero Chern number or trivial $\mathbb{Z}_2$ invariant and that the triplet phase is a chiral or helical gapped phase such as $p+ip$. Note that a $f+if$ phase on the square lattice is not gapped, since the diagonal line nodes of both $f$ waves overlap. Then there must be a topological phase transition, $\eta_c$, as a function of mixing ratio $\eta$. Of course, this transition might not be reached if $\eta$ varies between 0 and, say, 0.2 or 0.3; but suppose $\eta$ varies from small to large values then such a transition will occur. 
We note that from the list of possible singlet-triplet mixtures, Eq.\,\eqref{eq:mixed-states}, only the first case, $s+p$, represents a scenario where both phases are gapped: the $s$-wave corresponds to a topologically trivial phase but the helical $p$-wave is a fully gapped state with non-trivial $\mathbb{Z}_2$ topology. Note that a pure $s$-wave state is suppressed by the repulsive interaction, $U_{0}>0$; however, extended $s$-wave states can still appear. All other scenarios correspond to transitions from a gapless to a gapped state or between two gapless states. The former are possibly even more interesting: when a gapless and ``gapfull'', \ie a gap opening, term compete usually the latter wins easily. That suggests that indeed some of the mixed singlet-triplet states realize gapped, topologically non-trivial superonductivity.
Determining the exact values $\eta_c$ where such topological phase transitions occur and the investigation of phase diagrams as a function of $\eta$ will be reported elsewhere. 
Other lattices, \eg with hexagonal symmetry, might have a richer phenomenology since the $d$-wave representations are degenerate and thus prefer to form chiral singlet states of the type $d+id$ which are fully gapped.

In this work, we exclusively discussed the square lattice with its $D_4$ symmetry group. It contains five irreps, four of which are even (\ie the basis functions are even in momentum $k$) and one-dimensional.
As stated above, the form factor in helicity basis using time reversal partners, $\psi^{(T)}(k,\hel)$, transforms according to an even irrep of the point group. That raises the interesting question what would happen if $\psi^{(T)}(k,\hel)$ transforms according to an even irrep which is {\it two-dimensional}. Such a scenario is possible for the $D_6$ point group. The simplest example with $D_6$ symmetry is the triangular lattice.
%
%One interesting question, which can thus not be answered here, is which states would occure if $\psi^{(T)}(k,\hel)$ transforms according to an even 2D irrep, as it is possible, \eg in triangular lattice systems ($D_{6}$ point group). 
%
In analogy to Eq.\,\eqref{eq:mixed-states}, multiplying the even irreps with the  $\vec \gamma$ vector for the triangular lattice leads to the following singlet-triplet states:
\begin{equation}
%D_6:
%\hspace{-40pt}
\begin{split}
A_1 :&~ s + p\\
A_1 :&~ {\rm extended}\,s + p\\
A_2 :&~ i + h\\
E_2 :&~ d + \{f\,+\,p\}
\end{split}\end{equation}
The singlet contributions are $s$, $i$ and $d$. $s$ corresponds to isotropic, constant $s$-wave pairing, and extended $s$ to nearest-neighbor $s$-wave pairing. $i$ is an $i$-wave order with large angular momentum $\ell=6$ and $d$ refers to the two-component $d_{x^2-y^2}+id_{xy}$-wave order, which is chiral and topologically non-trivial (Chern number $C=2$). Triplet contributions $p$, $f$ and $h$ correspond to helical $p$-wave, $f$-wave and $h$-wave order. The latter has $\ell=5$. As for the square lattice, $p$-wave include the standard case but also ``extended'' $p$-wave with additional line nodes. For $E_2$, $\{f\,+\,p\}$ denotes a superposition of helical $p$- and $f$-wave order.

Instead of changing the lattice symmetry, also additional orbital or sublattice degrees of freedom might change the picture drastically, since the band index now incorporates multiple degrees of freedom and the structure of $\vec\gamma$ might be more complicated. This potentially leads to new types of superconducting states as well, which cannot appear for a one band model.

Finally, we briefly discuss the qualitative effects of disorder. To begin with, to treat random disorder is not a trivial issue and requires advanced methodology what is beyond the scope of this paper. Nonetheless, we wish to address a few points regarding disorder. For a superconducting state with a hard gap, small random disorder can be neglected (as long as the disorder strength is much smaller than the gap). Once the disorder strength becomes larger, it was recently shown that topological superconductors can undergo phase transitions either from trivial to topological phases or vice versa\,\cite{mascot-19prb235102,crawford-20prb174510}. Many of these transitions can be understood due to a renormalization of the band width and thus a rescaling of single-particle parameters. In addition, there might very well be many-body rescaling effects.

More complicated is the role of random disorder for ``gapless'' superconductors, \ie superconducting order parameters with nodal points or nodal lines. We expect that within the WCRG weak random disorder can be seen as a small perturbation. That is, if the leading and the second-leading instabilities are energetically close to each other (or even quasi-degenerate), weak disorder might easily cause a phase transition from one superconducting state to the other. In contrast, if leading and second-leading instabilities are well separated by a large energy gap, the WCRG results are likely to be robust.

%%%%%%%%%%%%%%%%%%%%%%%%%%%%%%%%%%%%%%%%%%%%%%%%%%%%%%%%%%%%%%%%%
%
%                                 C O N C L U S I O N
%
%%%%%%%%%%%%%%%%%%%%%%%%%%%%%%%%%%%%%%%%%%%%%%%%%%%%%%%%%%%%%%%%%
\section{Conclusion}
\label{sec:conclusion}
In this work, we have thoroughly investigated the square lattice Hubbard model in the presence of Rashba spin-orbit coupling.
We have developed an implementation of Rashba spin-orbit coupling into the weak-coupling renormalization group framework, which allows the study of non-centrosymmetric superconducting states with broken spin symmetry. These states feature mixing of singlet and (helical) triplet states, where smooth transitions between pure singlet and pure triplet states are possible as a function of any system parameter. While the mixing ratio can change continuously upon varying such parameters, we are still able to find the leading superconducting instability in an unbiased way. This is a major advantage of the weak-coupling renormalization group method. We also address the issue of how to properly deal with split energy bands within a ``Fermi surface method'', \ie a method which restricts paired electrons to momenta on the Fermi surface, and the transformation from helicity to spin basis.

\begin{acknowledgments}
We acknowledge instructive discussions with R.\ Thomale, P.\ Brydon, C.\ Honerkamp, M.\ Fink, and M.\ Klett. We further thank T.\ L.\ Schmidt for previous collaborations.
S.\ R.\ acknowledges support from the Australian Research Council through Grants No.\ FT180100211 and No.\ DP200101118.
This research was undertaken using the HPC facility Spartan hosted at the University of Melbourne.
\end{acknowledgments}

 \appendix
%%%%%%%%%%%%%%%%%%%%%%%%%%%%%%%%%%%%%%%%%%%%%%%%%%%%%%%%%%%%%%%%
%   
%                                 A P P E N D I X    A
%
%%%%%%%%%%%%%%%%%%%%%%%%%%%%%%%%%%%%%%%%%%%%%%%%%%%%%%%%%%%%%%%%
\section{Constraints on $\vec{\bs{\gamma}}$ due to symmetry}
\label{sec:appD}

%\subsection{Constraints on $\gamma$ due to symmetry}
%
The symmetry properties of the given lattice are reflected in properties of the Bloch matrix $\hat{h}$. We write the Bloch matrix as
\begin{equation}
\label{eq:bloch_general_multiorb}
 \hat{h}=\tau_{o}\otimes\gamma_{s}^{o}(k)\sigma^{s}\ .
\end{equation}
The orbital (sublattice) space parts, $\tau_{o}$, are elements of a complete set of $N\times N$ matrices, $\{\tau_{1},\tau_{2},\dots,\tau_{N^{2}}\}$, where $N$ is the number of orbitals per unit cell. Note that if $N=1$, which is the case for the system studied in this paper, $\tau_{1}\equiv1$.
The spin space part is given by the $k$-dependant functions $\gamma_{s}^{o}(k)$ and the Pauli matrices $\sigma_{s}$.

%%%%%%%%%%%%%%%%%%%%%%%%%%%%%%%%%%%%%%%%%%%%%%%%%%%%%%%%%%%%%%
%           F I G. 10
%%%%%%%%%%%%%%%%%%%%%%%%%%%%%%%%%%%%%%%%%%%%%%%%%%%%%%%%%%%%%%
\begin{figure*}[t]
 \includegraphics[width=1.6\columnwidth]{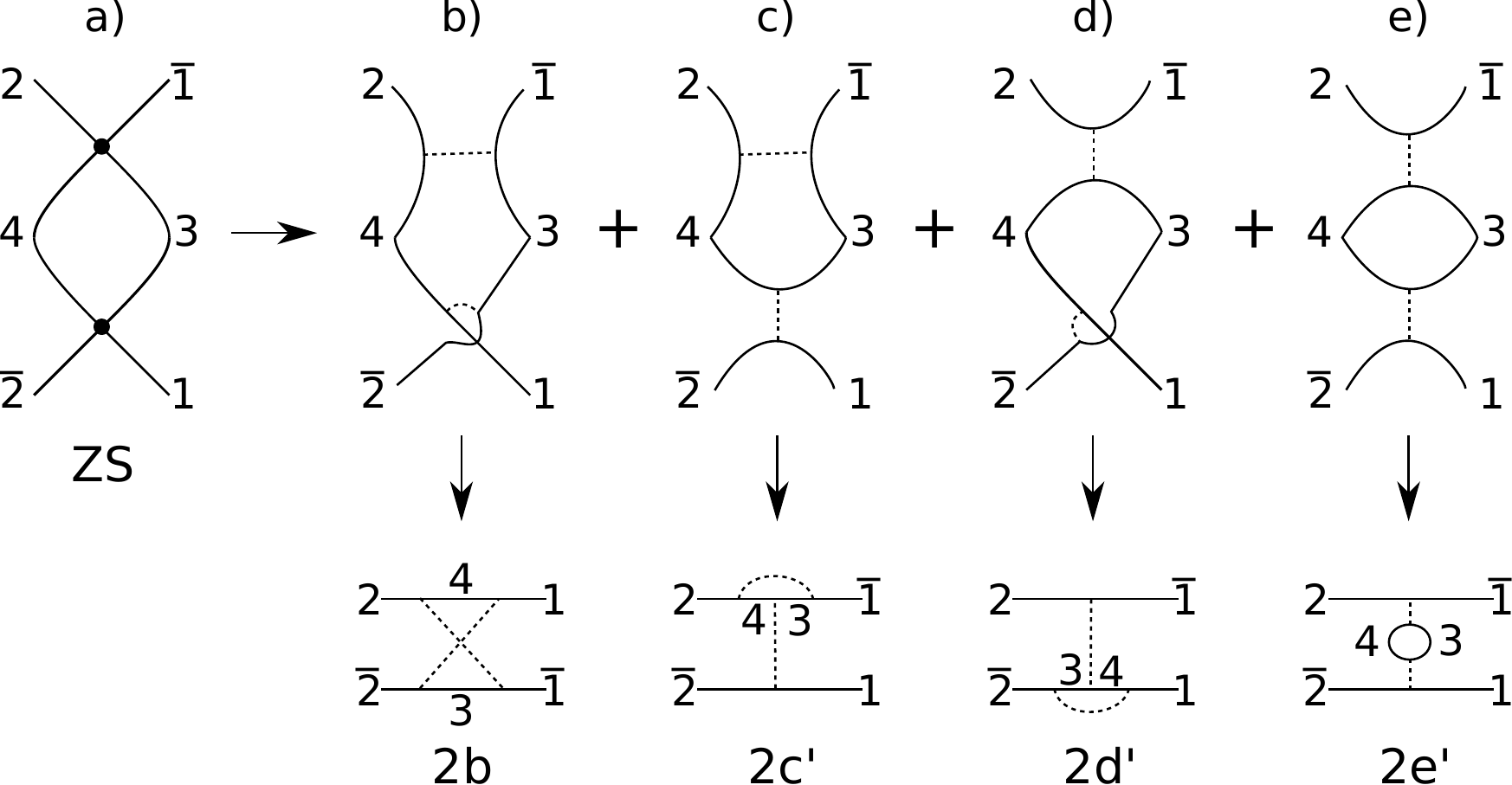}
 \caption{\label{fig:Feynman_trafo}Diagram ZS\,\cite{shankar_renormalization-group_1994}, where the interaction nodes are explicitly expanded in all possible configurations. The same can be done for the diagram ZS' by just switching $1$ and $\bar{1}$. Diagrams 2b, 2c$'$, 2d$'$, and 2e$'$ are in the convention of Refs.\,\onlinecite{raghu_superconductivity_2010,raghu_effects_2012,wolf_unconventional_2018}.}
\end{figure*}
%%%%%%%%%%%%%%%%%%%%%%%%%%%%%%%%%%%%%%%%%%%%%%%%%%%%%%%%%%%%%%

The symmetries of the system pose several constraints on $\gamma$, which are obtained by
\begin{equation}
 \hat{O}\hat{h}\hat{O}^{\dagger}=\hat{h}\ ,
\end{equation}
where $\hat{O}$ is the operator of the symmetry operation. Explicit representations of the relevant symmetries are listed in Tab.\,\ref{tab:symm_op}.
The resulting constraints on $\gamma$ due to these symmetries are listed in Tab.\,\ref{tab:symm_gamma} under the additional constraint of $\hat{h}$ being hermitian.

Note that for $N=1$ inversion and time reversal symmetry together set $\gamma_{x}=\gamma_{y}=\gamma_{z}=0$, whereas time reversal symmetry and the existence of a rotation center in the lattice for a rotation by the angle $\pi$ yields $\gamma_{z}=0$. If inversion symmetry and $\pi$-rotation symmetry is given, $\gamma_{x}=\gamma_{y}=0$, \ie spin rotation symmetry is also conserved.
\begin{table}[b]
 \caption{\label{tab:symm_op}List of symmetry operators, $\hat{O}=\hat{O}_{r}\otimes\hat{O}_{o}\otimes\hat{O}_{s}$, for several symmetries. $\hat{O}_{r}$ denotes the part of $\hat{O}$ which acts on real space, \ie on $\vec{r}$, $\hat{O}_{o}$ the part which acts on orbital (sublattice) space, and $\hat{O}_{s}$ the one which acts on spin space. Note that for time reversal, $\hat{T}=(\hat{T}_{r}\otimes\hat{T}_{o}\otimes\hat{T}_{s})K$, where $K$ denotes complex conjugtion. $\mathbbm{1}$ is the identity. For the systems studied here, $\hat{O}_{o}\equiv\mathbbm{1}$.}
 %%%%%%%%%%%%%%%%%%%%%%%%%%%%%%%%%%%%%%%%%%%%%%%%%%%%%%
 \begin{tabular}{c|c|c}
 \hline
 Symmetry   &   $\hat{O}_{r}$ & $\hat{O}_{s}$   \\[2pt]
 \hline\hline
 Time reversal  &
    $\mathbbm{1}$ & $-i\sigma_{y}$    \\[2pt]
 Inversion      &
    $-\mathbbm{1}$ & $i\sigma_{0}$  \\[2pt]
 Spin rotation around $z$   &
    $\mathbbm{1}$ & $\pm\begin{pmatrix}
                     e^{i\varphi/2} & 0 \\
                     0 & e^{-i\varphi/2}
                    \end{pmatrix}$   \\[10pt]
 $\pi$-rotation around $z$  &
    $\,{\rm diag}(-1,-1,1)\,$ & $\pm i\sigma_{z}$  \\[2pt]
    \hline
\end{tabular}
\end{table}
%%%%%%%%%%%%%%%%%%%%%%%%%%%%%%%%%%%%%%%%%%%%%%%%%%%%%%
%
%%%%%%%%%%%%%%%%%%%%%%%%%%%%%%%%%%%%%%%%%%%%%%%%%%%%%%
\begin{table}[t]
 \caption{\label{tab:symm_gamma}List of constraints on $\gamma$ due to symmetries.}
 \begin{tabular}{c|l}
 \hline
 Symmetry   &   ~Conditions   \\[2pt]
 \hline\hline
 Time reversal  &
    \,$\gamma_{0}(-k)=\gamma_{0}(k),$\\
    & \,$\gamma_{x,y,z}(-k)=-\gamma_{x,y,z}(k)$    \\[2pt]\hline
 Inversion      &
    \,$\gamma(-k)=\gamma(k)$  \\[2pt]\hline
 \,Spin rotation around $z$ \,  &
    \,$\gamma_{x}=\gamma_{y}=0$   \\[2pt]\hline
 \,$\pi$-rotation around $z$  &
    \,$\gamma_{0,z}(-k)=\gamma_{0,z}(k),$ \\
    & \,$\gamma_{x,y}(-k)=-\gamma_{x,y}(k)$  \\[2pt]
    \hline
\end{tabular}
\end{table}
%%%%%%%%%%%%%%%%%%%%%%%%%%%%%%%%%%%%%%%%%%%%%%%%%%%%%%

%%%%%%%%%%%%%%%%%%%%%%%%%%%%%%%%%%%%%%%%%%%%%%%%%%%%%%%%%%%%%%%%
%   
%                                 A P P E N D I X   B
%
%%%%%%%%%%%%%%%%%%%%%%%%%%%%%%%%%%%%%%%%%%%%%%%%%%%%%%%%%%%%%%%%
\section{Feynman diagrams}
\label{sec:appB}
%
%\subsection{Feynman diagrams}
%\label{subsec:Feynman}
%
In this section we explain why we use the Feynman diagrams given \eg in the review by Shankar\,\cite{shankar_renormalization-group_1994} (see Fig.\,\ref{fig:Feynman}), instead of the ones typically used in the WCRG papers\,\cite{raghu_superconductivity_2010,raghu_effects_2012,wolf_unconventional_2018} (see Fig.\,\ref{fig:Feynman_trafo}).
Both of the two variants are equivalent, as the ones in Fig.\,\ref{fig:Feynman} follow from the ones shown in Refs.\,\onlinecite{raghu_superconductivity_2010,raghu_effects_2012,wolf_unconventional_2018} taking into account Fermionic anticommutation relations. This is shown explicitly in Fig.\,\ref{fig:Feynman_trafo}, where the diagram ZS is expanded in the four diagrams of the particle hole channel of the same order ($U^{2}$), but different momenta (only the first one starts with $1$ on the top right leg, the other three with $\bar{1}$). Adding the same expansion of diagram ZS', which is obtained by simply swapping $1$ and $\bar{1}$, we obtain all scattering processes in the particle hole channel, including fermionic anticommutation of the incoming particles. This shows that
\begin{equation}
 \Gamma_{\rm ZS}+\Gamma_{\rm ZS'}\equiv\Gamma^{(2b)}+\Gamma^{(2c)}+\Gamma^{(2d)}+\Gamma^{(2e)}+1\leftrightarrow\bar{1}\ .
\end{equation}
The advantage of using the diagrams ZS and ZS' is that their numerical computation is more efficient, since we need to calculate only two intergrals (or a single one followed by the correct antisymmetrisation), whereas for the diagrams 2b-2e we would need to calculate four integrals, which ultimately yields the same result. Note that without spin-orbit coupling, the four diagrams 2b-2e can also be obtained by just a single integral; diagram 2a is not shown here\,\cite{raghu_superconductivity_2010,wolf_unconventional_2018}.

%%%%%%%%%%%%%%%%%%%%%%%%%%%%%%%%%%%%%%%%%%%%%%%%%%%%%%%%%%%%%%%%
%   
%                                 A P P E N D I X   C
%
%%%%%%%%%%%%%%%%%%%%%%%%%%%%%%%%%%%%%%%%%%%%%%%%%%%%%%%%%%%%%%%%
\section{Time reversal transformation of fermionic operators}
\label{sec:appC}
This section is essentially paraphrased from Appendix B in Ref.\,\onlinecite{sigrist_introduction_2009}. We add this here for the sake of completeness and to adapt the derivation to our notation.

%
%%%%%%%%%%%%%%%%%%%%%%%%%%%%%%%%%%%%%%%%%%%%%%%%%%%%%%%%%%%%%%
%           F I G. 11
%%%%%%%%%%%%%%%%%%%%%%%%%%%%%%%%%%%%%%%%%%%%%%%%%%%%%%%%%%%%%%
\begin{figure}[b!]
 \includegraphics[width=0.9\columnwidth]{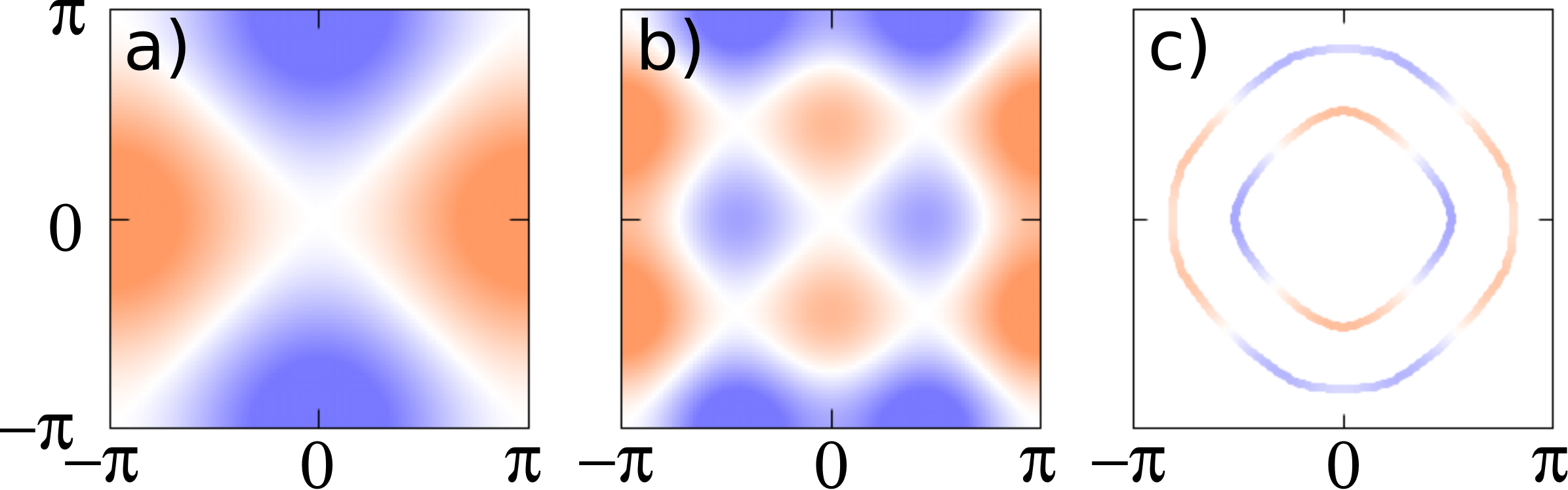}
 \caption{\label{fig:lh_B1}
Form factor plots for different lattice harmonics of the irreps of $B_{1}$ of the $D_{4}$ point group. a) Lowest order lattice harmonics, given by \eqref{fa}. b) Combination of two lattice harmonics, see \eqref{fb}, which produces an additional circular line node. c) Projection of b) onto an example set of spin-orbit split Fermi surfaces, such that the circular line node lies between them and is thus not visible.}
\end{figure}
%%%%%%%%%%%%%%%%%%%%%%%%%%%%%%%%%%%%%%%%%%%%%%%%%%%%%%%%%%%%%%
%
%
%%%%%%%%%%%%%%%%%%%%%%%%%%%%%%%%%%%%%%%%%%%%%%%%%%%%%%%%%%%%%%
%           F I G. 12
%%%%%%%%%%%%%%%%%%%%%%%%%%%%%%%%%%%%%%%%%%%%%%%%%%%%%%%%%%%%%%
\begin{figure*}[t]
 \includegraphics[width=1.8\columnwidth]{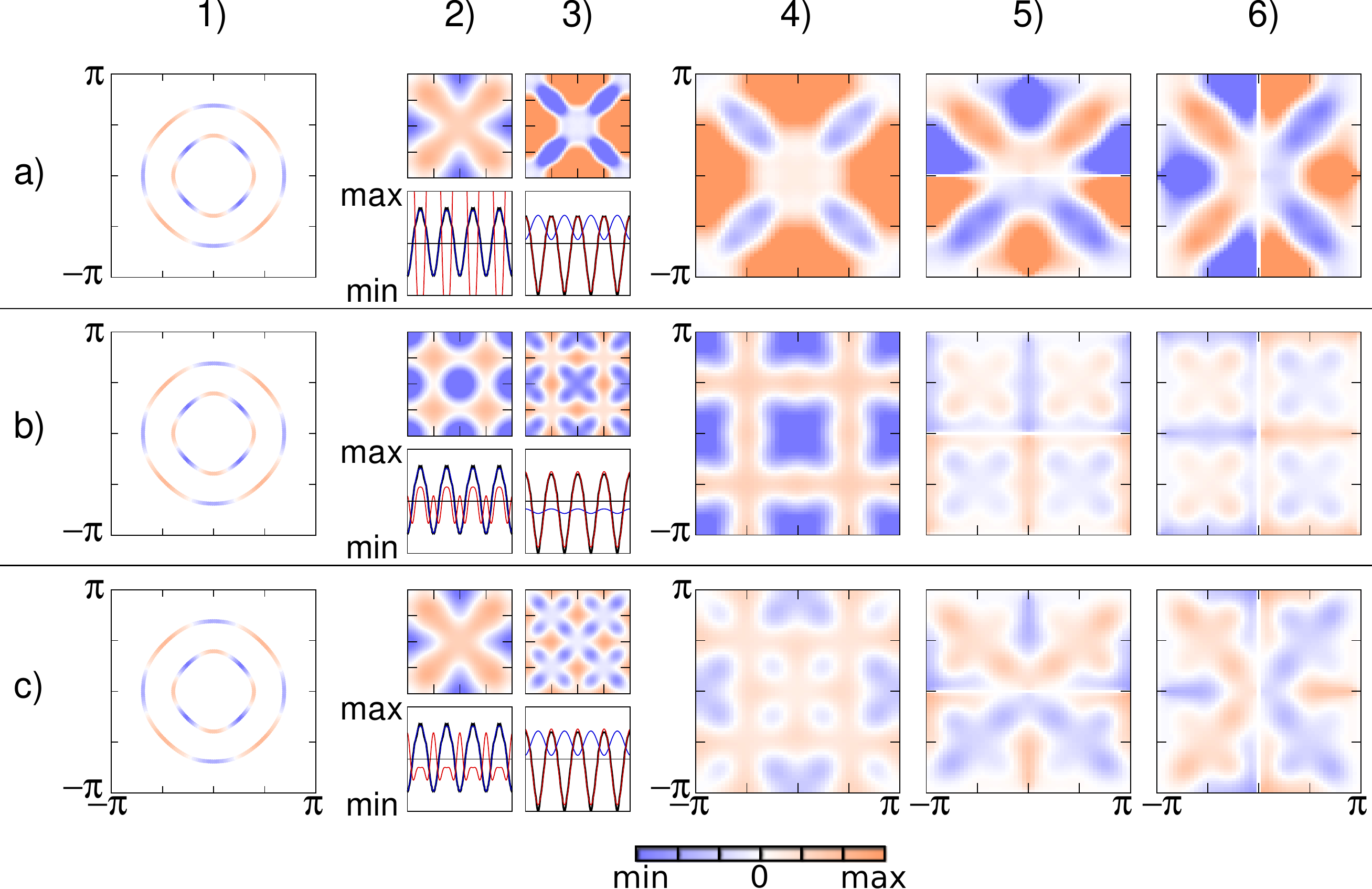}
 \caption{\label{fig:app_t_s_ambiguity}
Example of the ambiguity of fitting functions and the resulting mixing ratio between singlet and triplet states. Three different sets of fitting functions are shown as rows a), b), and c). Column 1): form factor of the superconducting order parameter in helicity basis, as given by the result of the WCRG calculations. Columns 2) and 3): Fitted functions to outer and inner Fermi surfaces. The blue line is the function fitted to the outer Fermi surface, the red line the one fitted to the inner Fermi surface. Columns 4) to 6): resulting functions for $d_{0}$, $d_{x}$, and $d_{y}$.
What we obtain are different extended $s$-waves mixed with different extended $p$-waves, \ie the symmetry of the $d$-vector components is the same in all three examples, but the nodal lines are very different. In particular, the mixing ratio, $\eta$, changes drastically between a), b), and c).
}
\end{figure*}
%%%%%%%%%%%%%%%%%%%%%%%%%%%%%%%%%%%%%%%%%%%%%%%%%%%%%%%%%%%%%%
%

The creation and annihilation operators in helicity basis, $b_{k\hel}^{\dagger}$ and $b_{k\hel}^{\pd}$, respectively, are given by unitary transformation of the corresponding operators in spin basis, \ie
\begin{align}
 b_{k\hel}^{\dagger}&=v_{k\hel}^{\up}c_{k\up}^{\dagger}+v_{k\hel}^{\dw}c_{k\dw}^{\dagger}\ ,\\[5pt]
 b_{k\hel}^{\pd}&=v_{k\hel}^{\up}c_{k\up}^{\pd}+v_{k\hel}^{\dw}c_{k\dw}^{\pd}\ .
\end{align}
Using the time reversal of the operators in spin space, which is given by
\begin{equation}
 \hat{T}\begin{pmatrix}
         c_{k\up}^{\dagger} \\
         c_{k\dw}^{\dagger} \\
        \end{pmatrix}
        =\begin{pmatrix}
          c_{-k\dw}^{\dagger} \\
          -c_{-k\up}^{\dagger} \\
         \end{pmatrix}\ ,
\end{equation}
yields the time reversal of the creation operator $b_{k\hel}^{\dagger}$ as
\begin{equation}
\label{eq:TR_b}
 \hat{T}b_{k\hel}^{\dagger}=\hat{T}\big(v_{k\hel}^{\up}c_{k\up}^{\dagger}+v_{k\hel}^{\dw}c_{k\dw}^{\dagger}\big)=v_{k\hel}^{\up *}c_{-k\dw}^{\dagger}-v_{k\hel}^{\dw *}c_{-k\up}^{\dagger}\ .
\end{equation} 
Comparing Eq.~\eqref{eq:TR_b} with 
\begin{equation}
 b_{-k\hel}^{\dagger}=v_{-k\hel}^{\up}c_{-k\up}^{\dagger}+v_{-k\hel}^{\dw}c_{-k\dw}^{\dagger}
\end{equation}
and using the relations $\evz_{\hel}(-k)=\evz_{-\hel}(k)$ and $e^{i\phi_{-k}}=-e^{i\phi_{k}}$, yields
\begin{align}
 b_{-k\hel}^{\dagger(T)}&\equiv\hat{T}b_{k\hel}^{\dagger}=-\hel e^{-i\phi_{k}}b_{-k\hel}^{\dagger}\ .
\end{align}
Thus, the phase factor $t(k,\hel)$ is given by
\begin{equation}
 t(k,\hel)=-\hel e^{-i\phi_{k}}
\end{equation}
for single orbital models.

%%%%%%%%%%%%%%%%%%%%%%%%%%%%%%%%%%%%%%%%%%%%%%%%%%%%%%%%%%%%%%%%
%   
%                                 A P P E N D I X   D
%
%%%%%%%%%%%%%%%%%%%%%%%%%%%%%%%%%%%%%%%%%%%%%%%%%%%%%%%%%%%%%%%%
\section{Transformation from helicity basis 
to spin basis}
\label{sec:appA}
Here we discuss how to perform the transformation of the form factor of the superconducting gap function from helicity basis to spin basis, which is explicitly given in Eqs.\,\eqref{eq:d0} and\,\eqref{eq:dv}. 
Of course, the transformation from helicity to spin basis is a simple unitary transformation; however, what is usually a simple task turns out to be a challenging problem for a Fermi surface method (such as WCRG), \ie the form factors $\psi^{(T)}(k\hel)$ are only given on the respective Fermi surface. 
For the transformation, however, we need the form factors at the same $k$-points for both bands, which we do not have. To solve this problem, we fit lattice harmonics with the same symmetry properties to each $\psi^{(T)}(k\hel)$ and thus obtain the form factors within the entire Brillouin zone.

However, since we fit a 2D function to a 1D manifold, the fit is underdetermined in the sense that we can produce arbitrary results far away from the Fermi surface, by including more and more lattice harmonics. This also means that, including an arbitrary number of arbitrarily high orders of lattice harmonics, we can always find a perfect fit with or without a line node exactly between the two Fermi surfaces, which enables us, in principle, to find a solution in spin space anywhere between pure singlet and pure triplet (note, however, that the irrep can not change). This becomes particularly easy when the splitting of the Fermi surfaces is large, \ie for large values of $\alpha_R$. An example to visualize this is shown in Fig.\,\ref{fig:lh_B1}. Here we see the lowest order lattice harmonic of the irreducible representation $B_{1}$, $f_{a}(k_{x},k_{y})$, in Fig.\,\ref{fig:lh_B1}\,(a), and a combination with a higher order lattice harmonic, $f_{b}(k_{x},k_{y})$, in Fig.\,\ref{fig:lh_B1}\,(b). The corresponding functions are given by
\begin{align}
\label{fa} f_{a}(k_{x},k_{y})=&\cos(k_{x})-\cos(k_{y})\ ,\\[5pt]
\label{fb} f_{b}(k_{x},k_{y})=&-\big[\cos(k_{x})-\cos(k_{y})\big]\\
          &+1.5\big[\cos(2k_{x})-\cos(2k_{y})\big]\ .\nonumber
\end{align}
Let us now imagine that we have a state similar to the one shown in Fig.\,\ref{fig:lh_B1}\,(c), \ie where we have two Fermi surfaces separated such that the circular line node in Fig.\,\ref{fig:lh_B1}\,(b) lies between them, and where the superconducting order parameter has opposing signs on each Fermi surface. Then we could either use the function of Fig.\,\ref{fig:lh_B1}\,(a) for fitting, which would yield a dominant triplet state, or we could use the one of Fig.\,\ref{fig:lh_B1}\,(b), which would yield a dominant singlet state. This is the problem of ambiguity resulting from fitting a 2D function to a 1D manifold.

However, from a physical perspective, we implement a few criteria for the fitting process to avoid ambiguities as described above; indeed these criteria heavily limit the amount of freedom for the resulting ratio of singlet and triplet pairing. These are given in the following, sorted by decreasing importance:
\begin{enumerate}
 \item The form factor should not be much larger far away from the Fermi surface than on the Fermi surface itself.
 \item We use lower order lattice harmonics rather than higher orders. It is often necessary to find a compromise between low order and low fitting error.
 \item We limit the number of lattice harmonics used for the fit to a maximum of three, and using fewer where possible.
\end{enumerate}
In the example presented in Fig.\,\ref{fig:lh_B1} above, we would thus choose to use the function of Fig.\,\ref{fig:lh_B1}\,(a) to fit and obtain a mainly triplet state, since it uses lower order harmonics than the function of Fig.\,\ref{fig:lh_B1}\,(b).

Fig.\,\ref{fig:app_t_s_ambiguity} shows an example from our results, where the form factor transforms according to the irrep $A_{1}$. We see that we can fit different functions equally well to the Fermi surface, which yield drastically different results.

%\sw{Comment: These ambiguities can appear everywhere, they are just more difficult to handle when the splitting is large. In other words: You have to make some serious effort to get different results for a small splitting and have a good fit at the same time.}

The main results of this paper are, however, not affected. Firstly, the irrep always remains the same, regardless of the fitting process; only the ratio between singlet and triplet can be affected by fitting ambiguities. Secondly, only for large SOC, \ie significant splitting of the Fermi surfaces, it becomes more challenging to handle such ambiguities. Thirdly, when sweeping through entire parameter ranges (such as phase diagrams) it turns out that a single point might still allow for such fitting ambiguities, but parameter points close by do not. By adiabaticity we can thus find further evidence in favor of one of the fitting options.
We stress again that the issue of fitting ambiguities is relevant to any Fermi surface method and not specific to WCRG.

\bibliography{wcrg2}

\end{document}